\documentclass[11pt,fleqn,a4paper]{article}

\usepackage{lineno}
\usepackage{appendix}
\usepackage{amsmath}
\usepackage{setspace}
\usepackage{ccaption}
\usepackage{latexsym}
\usepackage{appendix}
\usepackage{natbib}
\usepackage[FIGTOPCAP]{subfigure}
\usepackage[none]{hyphenat}
\usepackage{anysize}
\marginsize{2.5 cm}{2.5 cm}{2 cm}{2.5 cm}
\usepackage{rotating}
\usepackage{url}
\usepackage{dcolumn,longtable}
\usepackage{caption}
\usepackage{pslatex}

\begin{document}

\onehalfspacing


\title{Structural distortions in the Euro interbank market:\\ The role of `key players' during the recent market turmoil}

\author{Caterina Liberati, Massimiliano Marzo, Paolo Zagaglia and Paola Zappa\footnote{Liberati: Economics Department, Universit\`{a} degli Studi Milano-Bicocca; caterina.liberati@unimib.it. Marzo: Department of Economics, Universit\`{a} di Bologna; massimiliano.marzo@unibo.it. Zagaglia: Department of Economics, Universit\`{a} da Bologna; paolo.zagaglia@unibo.it. Zappa: Economics Department, Universit\`{a} degli Studi Milano-Bicocca; paola.zappa1@unimib.it.} }

\date{{\small This version: \today}}

\maketitle

\thispagestyle{empty}

\setcounter{footnote}{0}

\renewcommand{\thefootnote}{\arabic{footnote}}

\singlespacing

\begin{abstract}

We study the frictions in the patterns of trades in the Euro money market. We characterize the structure of lending relations during the period of recent financial turmoil. We use network-topology method on data from overnight transactions in the Electronic Market for Interbank Deposits (e-Mid) to investigate on two main issues. First, we characterize the division of roles between borrowers and lenders in long-run relations by providing evidence on network formation at a yearly frequency. Second, we identify the `key players' in the marketplace and study their behaviour. Key players are `locally-central banks' within a network that lend (or borrow) large volumes to (from) several counterparties, while borrowing (or lending) small volumes from (to) a small number of institutions. Our results are twofold. We show that the aggregate trading patterns in e-Mid are characterized by largely asymmetric relations. This implies a clear division of roles between lenders and borrowers. Second, the key players do not exploit their position of network leaders by imposing opportunistic pricing policies. We find that only a fraction of the networks composed by big players are characterized by interest rates that are statistically different from the average market rate throughout the turmoil period. \\ \\
{Keywords:} Market microstructure, network analysis, money markets, asset pricing. \\
{JEL classification:} D85, G01, G10, G21.

\end{abstract}

\newpage

\singlespacing

\begin{quote}
\textit{``In the no-holds barred world of trading over-the-counter derivatives in the interbank market, traders and brokers view themselves as combatants in a professional market, where you lose one day, but can win the next. (...) The industry is reluctant to fully automate OTC trading because it would result in a more open and transparent market and erode the informational advantages of the big dealers. Smaller banks have little choice but to abide by the rules.''}

\hspace{3.5in}\citet{mackenzie}
\end{quote}

\onehalfspacing

\section{Introduction}

Transactions in the money market represent one of the key funding vehicles for financial institutions. The structure of the market contemplates a primary provider of liquidity, namely the central bank. A large part of the market, though, consists in banks lending to each other. The centrality of the money market has been remarked during the recent financial turmoil, which started out in August 2007 with a freeze in interbank lending. The recent experience has showed that a lock-up in interbank activity carries systemic implications for the markets for all the other assets \citep[see][]{holthausen}. Therefore, understanding the patterns of trading in the interbank market is crucial to evaluating the properties of its functioning both during normal conditions, and during times of stress.

The literature on networks in the interbank market has grown large over the last few years. A large part of the available studies focus on describing the behaviour of banks on a daily frequency. Several papers use methods from network topology to study the types of linkages between banks \citep[e.g., see][]{Bech:2008,iazzetta}, and what role these relations play over the short run \citep[e.g.,see][]{bech:2}. Recently there has also been a focus to characterize the systemic risk of contagion of a liquidity squeeze, namely the risk that a shortage of funding for a bank may generate adverse effects for the entire financial system \citep[e.g., see][]{drehmann}.

This paper takes a long-term view on the functioning of the Euro overnight money market. We identify the structure of relations over a yearly frequency in the Electronic Market for Interbank Deposits (e-Mid), the main electronic platform for unsecured lending. Our analysis covers the period of the recent financial turmoil between 2006 and 2009. Hence, we focus on the changes in the patterns of exchange that have arisen from the disruption of standard market activities in the Euro area after 2007. 

The platform e-Mid provides a transparent and non-anonymous market. The organization of exchanges allows market participants to differentiate with regard to counterparty characteristics, and to evaluate the trading behavior of an institution. This suggests that the reputation of a bank is a key factor in the establishment and maintenance of lending relations, especially over the long run\footnote{The role of market reputation is also stressed by \citet{idier} and \citet{zagaglia} in the context of the over-the-counter segments of the interbank market.}. 

Our results indicate that the aggregate trading patterns in e-Mid are characterized by largely asymmetric relations in each network. These imply a clear division of roles between lenders and borrowers. To put it more clearly, there are institutions that exercise a leading control both on the lending and the borrowing side of the market. Thus, the subsequent question of interest has to do with the identification of these banks.

We study the behavior of the `key players' in e-Mid. These are institutions that lend (borrow) large amounts to many counterparties, while borrowing (lending) small amounts of liquidity from few banks of the system. The key players can be thought of as `locally-central banks' in the context of their network of lending relations. From an intuitive viewpoint, these banks are the driving forces of the demand and supply side of the market.  Understanding the behavior of the key players provides insights on the bottlenecks in the distribution of liquidity and on their role in the changing market structure of e-Mid.

The presence of key players matters especially in the context of networks with an asymmetric structure. The reason is that they generate a bottleneck problem in the distribution of liquidity. In this case, we can consider the hypothesis that stable relations are the expression of the `market power' of a central node\footnote{The issue of market power in financial networks is also investigated by \citet{kraenzlin}, who study the price setting behavior in the Swiss Franc repo market during the turmoil period. They find that banks use both their market power and private information to offer different lending rates depending on the characteristics of their counterparties.}. We can interpret this as a friction in the distribution of liquidity across the banking system. The relevant question of interest is whether the presence of this friction systematically benefits the key players. In other words, we would like to study whether the key players exploit their leading position by demanding higher (lower) lending (borrowing) interest rates than the market average.

The analysis of the key players sheds light on two main empirical facts. We find that the composition of the group of key players - both for the supply and the demand side - has changed every year since 2006. This indicates that e-Mid is not composed by market players with medium- or long-term trading strategies. Rather, banks' patterns of exchange appear contingent on short-term developments. In terms of pricing policies, the key players do not exploit their market position. For instance, on the lending side, only a fraction of the networks composed by big players charge interest rates that are statistically higher than the average market rate throughout the sample period. 

The results presented in this paper are relevant for understanding how the European Central Bank (ECB) can control the monetary transmission through the interbank market. In particular, if a group of banks control the distribution of liquidity to the financial system, they may distort the conditions of supply to their counterparties. The effects of these distortions may play against the stance of liquidity supply implemented by the ECB. This issue is especially relevant in the context of the recent market turmoil. In fact, during this period, the ECB put in place a number of extraordinary liquidity-supply operations with the aim of easing tensions in the market \citep[see][]{lenza}.

This paper is organized as follows. In Section 2, we discuss some empirical findings from the available literature and provide theoretical considerations on the focus of our paper. Section 3 considers the information available in our dataset of the e-Mid market. Section 4 provides a descriptive analysis of the data. Section 5 describes our approach to network analysis. Section 6 presents the main patterns of the networks that provide the ground for our study of key players. Section 7 discusses our methodology for identifying the largest net liquidity providers or borrowers in each network. In Section 8, we discuss the characteristics of the key players. Section 9 proposes some concluding remarks.

\section{Previous empirical findings and theoretical considerations} 

The role of `centers' in the distribution of liquidity is investigated by \citet{craig}. They use data on German banks to show that these institutions do not lend directly to each other. They tend to supply liquidity through the intermediation of locally-central banks. The centrality of a bank is largely correlated to its idiosyncratic characteristics. For instance, banks with large balance sheets are associated with a central position in the network. Overall, these findings suggest the idea that the money market may be characterized by `key players' with a determinant influence on the aggregate liquidity imbalance. 

The literature has also investigated the pricing impact of `relationship lending' in the interbank market. \citet{cocco} find that, within their network of relations, banks pay lower borrowing rates and trade with counterparties that face uncorrelated liquidity shocks. \citet{fecht} provides a test on the role of stable connections in the German money market. Their empirical results indicate that banks operating in formal networks do not enjoy any preferential price treatment. Rather than the existence of a relationship, the key factor is represented by the ability of an institution to avoid a net `squeeze' \textit{vis-\`{a}-vis} the aggregate liquidity supply. 

We should stress that the organization of the Euro money market is affected by the institutional features that govern the primary supply of liquidity by the European Central Bank. In particular, these features provide the scope for a limited set of banks to play a leading role in the interbank market. \citet{idier} and \citet{zagaglia} point out that the rules for taking part to the ECB tenders have a discriminatory character. As a consequence, they can generate systematic distortions in the distribution of liquidity to the banking system. Private banks can take part to the liquidity operations only if they are listed as eligible counterparties by the ECB. In addition, banks face relevant administrative costs for taking part to the ECB tenders. These transaction costs can represent a disincentive especially for banks of smaller dimensions. 

Interestingly \citet{idier} provide empirical evidence suggesting that trading in the overnight segment of the uncollateralized interbank market is affected by asymmetric information among counterparties. \citet{zagaglia} extends this result to the term segments of the Euro money market. Overall, these studies suggest that there are banks that are not constrained by the needs of compulsory reserve management. These institutions engage in significant trading activities with the rest of the interbank market. They can, thus, collect and exploit information that smaller banks do not have about the aggregate liquidity imbalance. In other words, these studies hint at a relation between information asymmetry between banks and the behaviour of a key player in the interbank market. 

The relation between network structure and market information is discussed also by \citet{brunetti:1} for the case of S\&P 500 futures contracts. They find that indicators of network organization are useful for understanding changes in financial market variables, as they predict traded volume and intradaily duration. Hence, the properties of a network can be interpreted as a mere metrics for information flow. 

What kind of `information' does the literature point to? \citet{babus} charaterize the information leading to network formation as information about counterparty risk, or `risk of contagion'. They provide a theoretical model suggesting that banks minimize the trade-off between costs and benefits of creating a network by choosing for partners that are resilient to contagion from adverse shocks. Thus, an equilibrium network carries a probability of contagion equal to zero. In addition, the banks outside the network face credit rationing from part of the network components.

\section{The dataset}

We investigate the Electronic Market for Interbank Deposits (e-Mid) repository, an electronic interbank platform fully centralized and operating in Milan. e-Mid is a unique market in the Euro area and records all the transactions occurred in the market among the registered banks.
 
This market has several peculiar characteristics. The first one is that it consists of two sub-markets, and consequently of two types of transactions, which follow different rules. In the 'ask' (or buy) market the transaction is started by the borrower - i.e., the aggressor -, who buys liquidity from the lender - i.e. the quoter. By contrast, in the 'bid' (or sell) market the transaction is initiated by the lender - i.e., the aggressor, who sells liquidity to the borrower, i.e. the quoter. Transactions in e-Mid involve money exchange on a different maturity structure, from overnight to eleven-month length. However, the most part of trades occurs on overnight maturity contracts. Banks can choose their trading counterparty, whereas the information on rates and amounts is made public. In addition, the minimum trade size is established \textit{a priori}. 

The information available makes a distinction between regular size with a minimum amount for transactions of 1.5 million euros, and a large size with a minimum amount of 100 million euros. For each transaction executed throughout the system, a record is produced which provides information on the identity of the aggressor and the quoter, the amount traded, the interest rate, the date and time of delivery, the loan length and type. Because of privacy concerns, the identity of each e-Mid member is represented by a unique six digit code, the first two being the country of origin and the following four a 0001 to $nnnn$ code. This allows to determine the nationality of a bank, but not its identity.

\section{Empirical patterns on a yearly frequency}  

Our sample counts 305,489 overnight `ask' and `bid' instances which took place from January 1 2006 to December 31 2009. Since the overnight transactions represent the overwhelming majority of the interbank market, we exclude the trades with a longer duration from the sample. 

\begin{figure}
 \centering
 \caption{Descriptive statistics for interest rates on recorded transactions}
\includegraphics[width=0.7\textwidth]{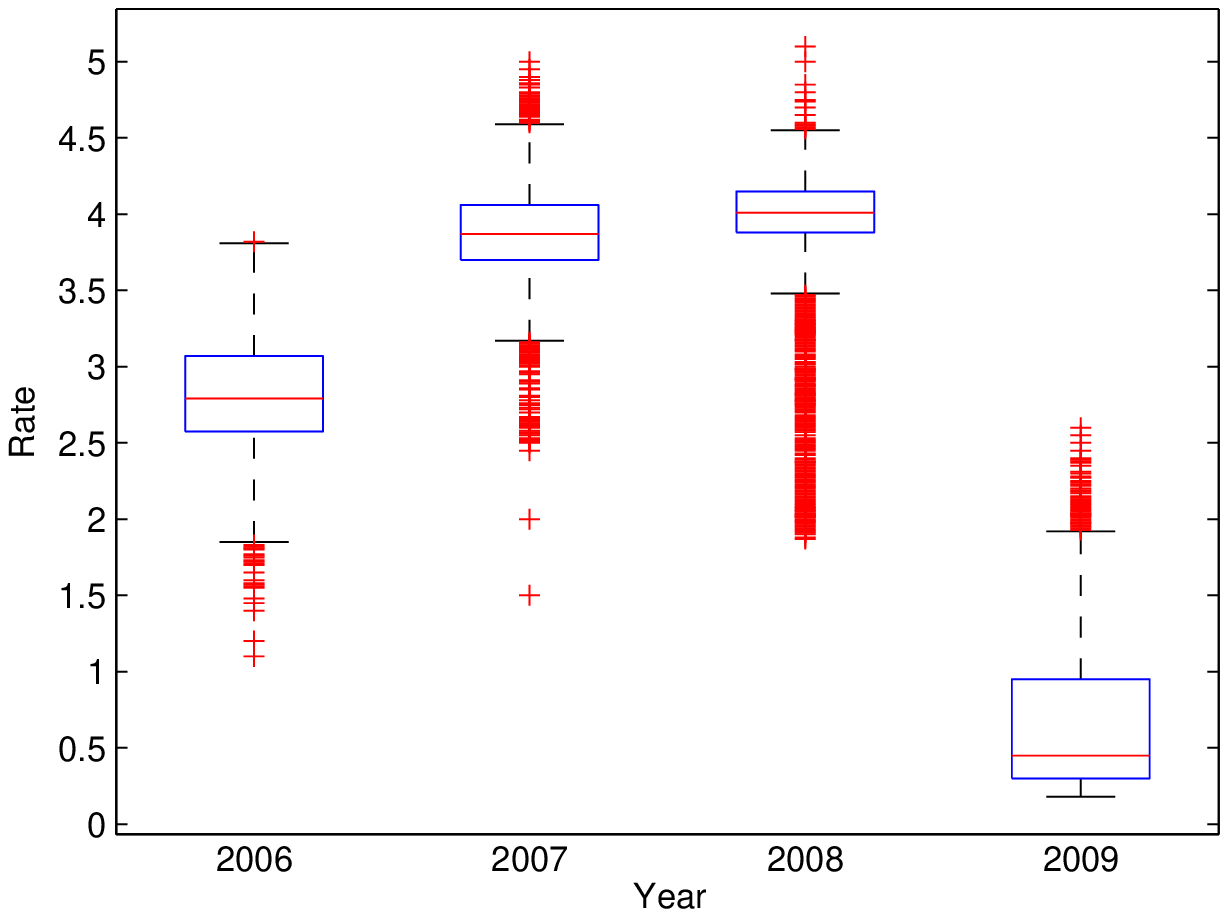}
\label{Rate}
\end{figure}

The impact of the 2007 market turmoil on the transaction patterns in e-Mid has several dimensions. Table \ref{tab:tab1} reports the number of transactions and the number of market participants. Both these figures have decreased over time\footnote{There are banks that join or leave the system at different times. Across the sample period, the total number of actors operating in e-Mid is 194.}. The size of trades has also changed noticeably after the turmoil. This is shown in Table \ref{Amount}. The freeze in market activity that characterizes the tumoil period is reflected by the fall of the total volume traded (Sum), which decreases by 14\% from 2006 to 2007, undergoes a further reduction of 20\% from 2007 to 2008, up to decay by 40\% through 2008 to 2009. We observe a similar trend for the average amount (Mean), which drops by 30\% from 24.655 to 17.391. Similiarly, this tendency has affected the evolution of the lending rates over the 2006-2009 period. Figure \ref{Rate} reports the distributions of the rates for each sample year. Both the mean and the median rates have risen in 2007 and 2008 relative to 2006. The market has however experienced a consistent drop of lending rates in 2009. 

The changes in the heterogeneous behavior of the e-Mid participants deserve attention. Table \ref{bank} reports some descriptive statistics on volumes lent and borrowed per bank. In each year, only a fraction of the banks active in e-Mid operates on both sides of the market. The share of lenders never exceeds 93\% of the total number of banks. This trend is constant over time. For the borrowing side, the figure drops from 89\% in 2006 to 84\% in 2009.   

These considerations suggest that e-Mid is a market where banks with dominant roles can co-exist. There are financial institutions that focus on lending activities, and banks devoted mostly to borrowing. Additional evidence on this point can be obtained by observing standard deviations of the distributions of deposits (both lent and borrowed) per bank. The high dispersion of the data with respect to the average, in fact, indicates the presence of rather different approaches to trading. 

The dynamics of the distributions of volume traded reveals a high concentration both on the lending and on the borrowing side. In 2006, ten banks - all of which Italian - realize the 39\% of the total volume lent, a percentage that drops to 34\% in 2007 and that reaches 24\% in 2008 and 29\% in 2009. This indicates a trend of greater sharing of liquidity across the market. A similar picture emerges for the demand for deposits. The level of resource concentration is stronger among borrowers than among lenders. The first ten banks - again, all of them have Italian nationality - account for 56\% of the market in 2006, with a share declining to 39\% in 2009\footnote{We should also add that, as accounted for by other studies \citep[e.g., see][]{ecb}, nationality of market participants does play a key role in e-Mid. The counterparties belong to 16 different countries, with Italian banks being the most important ones. The weight of Italian banks in e-Mid increases over time, as as an increasing number of foreign banks leaves the market. There is division of roles across nationalities, with French, Greek and Dutch banks mostly lending, and British banks mostly borrowing.}.

\section{Methodology for network analysis}

In order to analyze the structure of the money market and to detect individual behavior, we use methods from the so-called Social Network Analysis - SNA hereafter - \citep[e.g., see][]{Wasserman:1994, Scott:2005, Borgatti:2009}. This framework has recently been applied to different contexts ranging from the study of interpersonal relations to interorganizational dynamics. With regard to the relationships among banks, SNA has proven effective in examining the topological properties of the money market \citep{Masi:2006, Iori:2008}. In this paper, we apply the tools of SNA with the final aims of studying the stability of network relations and of identifying the most important actors.

Building on previous studies \citep[e.g., see][]{Iori:2008}, the transactions are represented as a network $N(V,E)$. The network nodes $V$ are the banks, and ties $E$ are identified as the money lent from one institution to another. To this purpose, aggressors and quoters are reclassified as lenders and receivers (or the opposite, depending on the sub-market explored), following the money flow direction. Therefore, we assume that a tie exists from the bank $i$ to the bank $j$ when $i$ lends money to $j$, independently from the origin of the transaction (i.e. the transaction is of 'ask' or 'bid' type). We have the following two cases:
\begin{enumerate}
\item for bid transactions, the tie goes from the aggressor (lender) to the quoter (receiver)
\item for ask transactions, the tie goes from the quoter (lender) to the aggressor (receiver).
\end{enumerate}

\begin{figure}[!p]
 \centering
 \caption{Network of transactions (2006-2009)}
 \label{fig:net}
 \subfigure[2006\label{fig:net1}]
   {\includegraphics[width=0.42\textwidth]{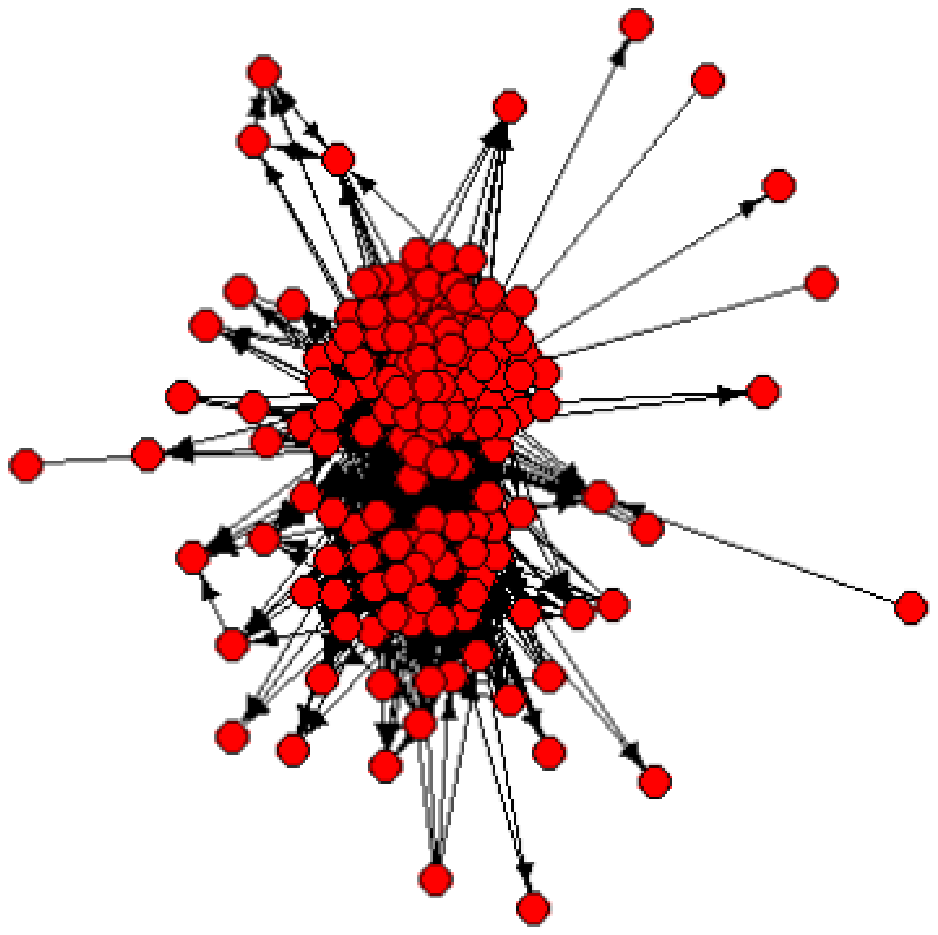}}
 \subfigure[2007\label{fig:net2}]
   {\includegraphics[width=0.42\textwidth]{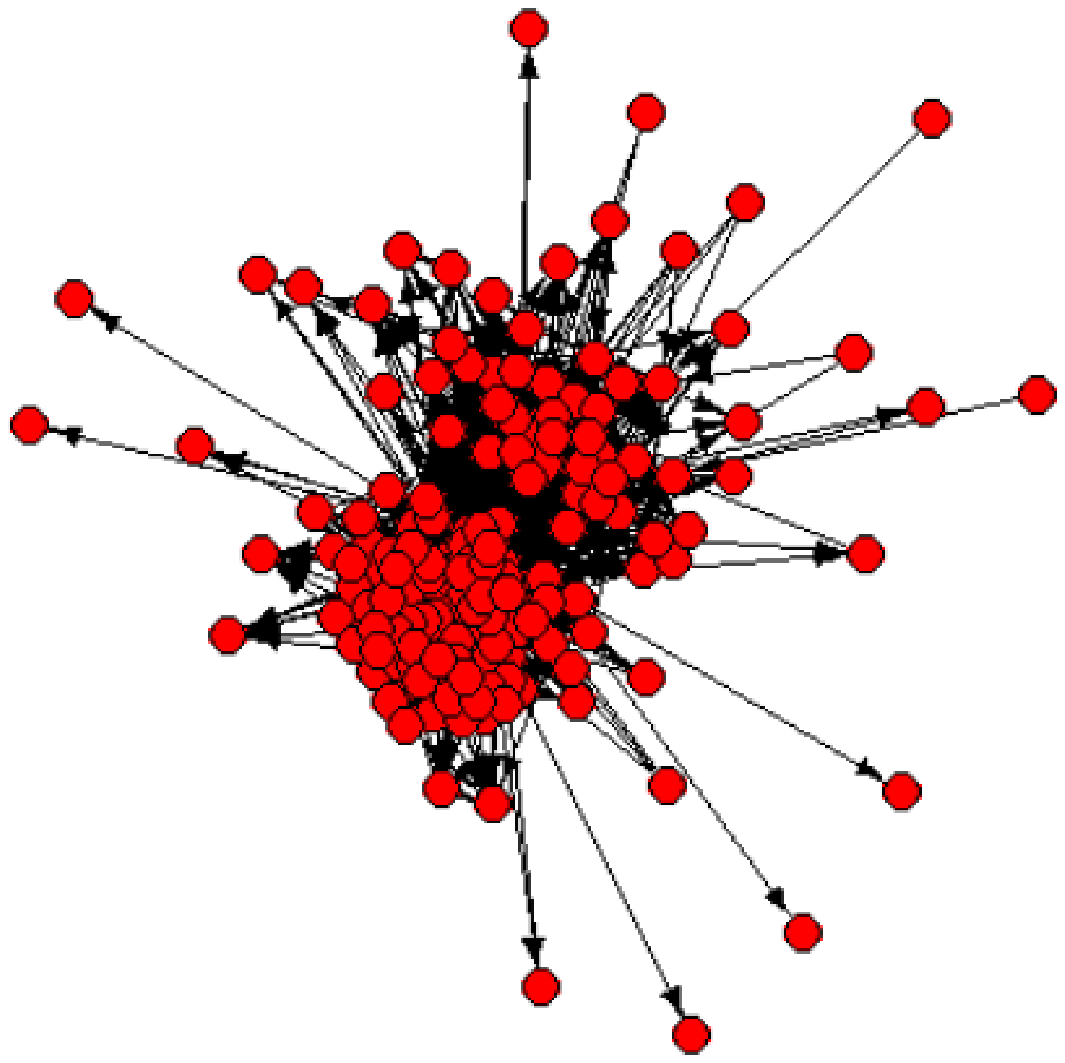}}
   \\
 \subfigure[2008\label{fig:net3}]
   {\includegraphics[width=0.42\textwidth]{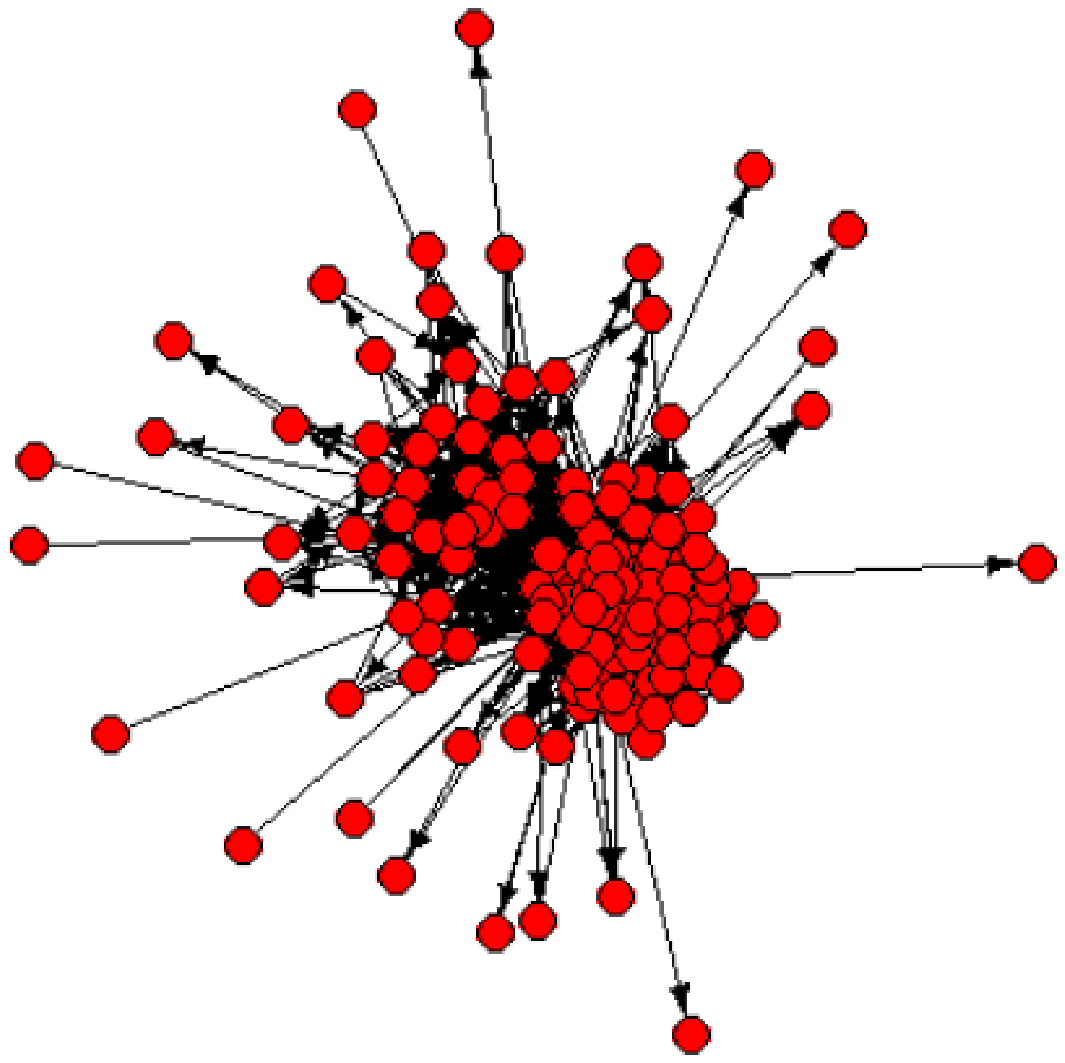}}
 \subfigure[2009\label{fig:net4}]
   {\includegraphics[width=0.42\textwidth]{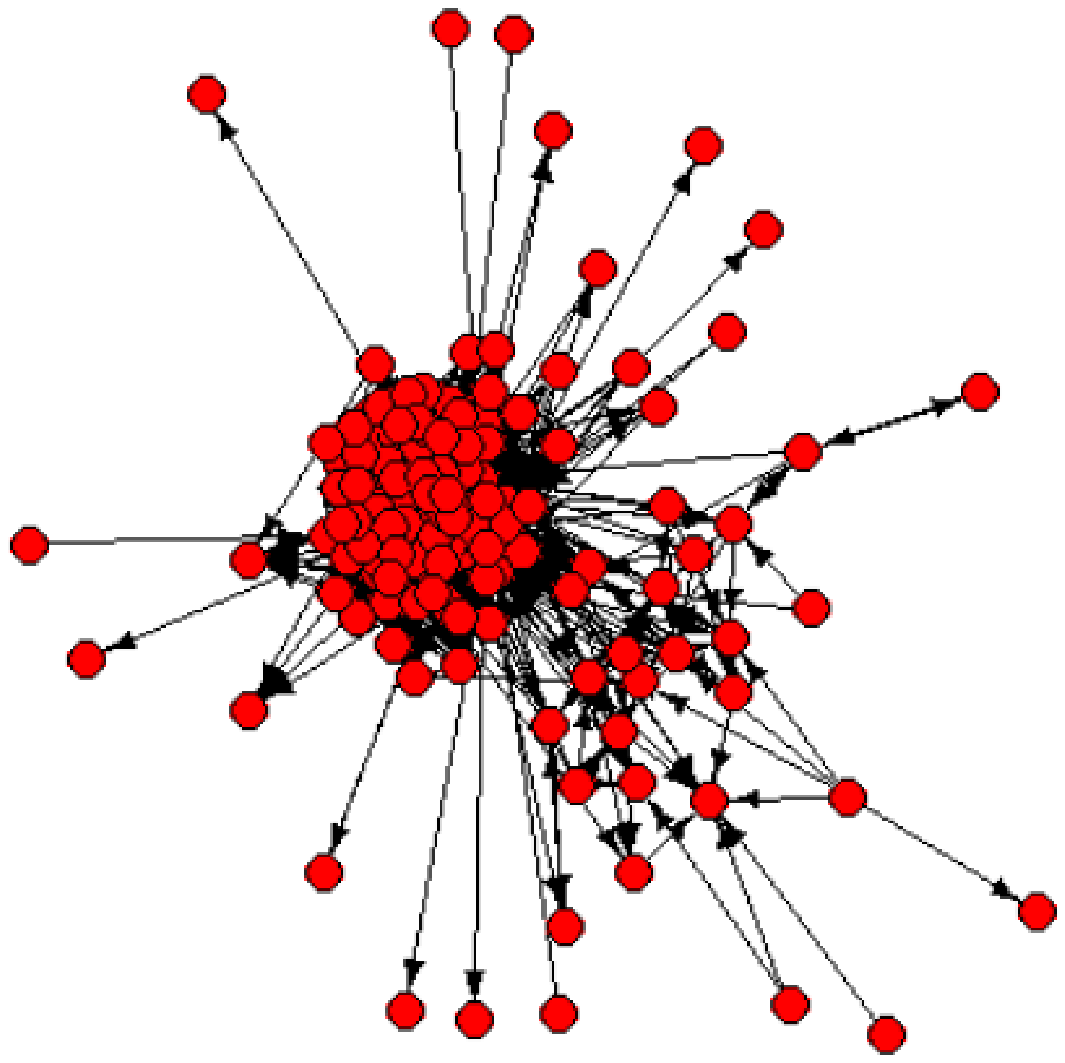}}
\end{figure}

Since the tie from $i$ to $j$ ($e_{ij}$) is different from the tie from $j$ to $i$ ($e_{ji}$), the network is defined as `directed'. To each tie, we attach a weight $w_{ij}$ that represents the amount of money that $i$ lends to $j$ over a given time span. The most available studies build and examine the evolution of a network at a daily frequency, in order to detect how money exchanges take place in the short term. This paper has a different aim: we are interested in identifying the market structure, i.e. existence of stable relationships among pairs or subgroups of banks, and the persistence of roles and market positions. Finally, we want to interpret our results in the light of the turmoil. Therefore, it seems reasonable to consider a long time span, i.e. one solar year. The weight $w_{ij}$ is computed as the annual amount lent from $i$ to $j$. Formally:
\begin{equation}\label{eq:app1}
w_{ij} = \sum^{H}_{z=1}{t_{ijz}}
\end{equation}
where $t_{ijz}$ is the amount of the transaction $z$ from $i$ to $j$ per year, and $H$ denotes the number of transactions from $i$ to $j$ per year. Then, we build four networks, one for each year. These are displayed in Figure \ref{fig:net}. Each network corresponds to an asymmetric adjacency matrix $W$ of size $n\times n$ whose generic element is $w_{ij}$ ($i$=1,...,n;$j$=1,...,n;$i\neq j$).

The analysis is performed on two levels. First, we consider the full structure of transactions at the network level for each year. This provides some general insights on the interaction among banks. We then shift our focus to the role of each bank in the network. This second level of analysis leads to the identification of market inefficiencies or bottlenecks, i.e. existence and persistence over time of banks that control the supply or demand of deposits in the market. Their behavior and pricing policies are then studied in detail. We repeat this process for the four yearly networks. This provides insight on the time variation of the stability of long-term network reations over time\footnote{To run the analysis, we use the software packages sna \citep{Butts:2008, Butts:2010}, igraph \citep{Csardi:2006} and tnet \citep{Opsahl:2011}, developed within the R statistical computing environment and specifically designed for network studies.}.

\section{Network relations over the long run}

\subsection{Empirical results from network-level analysis}

We start by computing the so-called `network density'. The density of a directed network or graph is the proportion of possible arcs (or ties) that are actually present in the network. This is, a measure of completeness. Formally, the density $\Delta$ is the ratio of the number of arcs present $L$ to the maximum possible \citep[see][]{Wasserman:1994}:
\begin{equation}\label{eq:app2}
\Delta = \frac{L}{n(n-1)}
\end{equation}
Since an arc can be seen as an ordered pair of nodes, there are $n(n-1)$ possible arcs. The density of a directed network takes values between 0 - if no arcs are present - to 1 - if all arcs are present. In our network the value of the density is moderate and almost constant over time. In fact, the density coefficient is 0.17 in 2006 and 2007, 0.16 in 2008 and 0.15 in 2009. Overall, indicates that less than 20\% of the possible links among the banks are put in place in the network.

Reciprocity is defined at dyadic level, in the sense that it focuses on the relationship between a pair of nodes. This concepts identifies a mutual exchange of money within a pair of banks. Therefore, it implies the existence of a non-hierarchical relationship among them. A node pair ($i$,$j$) is called `reciprocal' if there are arcs between them in both directions. Hence, the reciprocity of a directed graph is the proportion of all possible ($i$,$j$) pairs which are reciprocal, provided there is at least one arc between $i$ and $j$. Like the density, the reciprocity index varies between 0 and 1.
\begin{equation}\label{eq:app3}
r = \frac{\sum_{ij}e_{ij}e_{ji}}{L}
\end{equation}
The values of the reciprocity index for e-Mid are very low and suggest the lack of bidirectional exchange between banks. The percentage of reciprocal dyads is always lower than 30\% and declines significantly over time. It is equal to 28.01\% in 2006, 27.73\% in 2007, 24\% in 2008 and 16\% in 2009. This indicates that most of the relationships are asymmetric. In other words, there seem to be a strong distinction of roles in e-Mid throughout the observation period.

\subsection{Perspectives from actor-level analysis}

In the second level of the analysis, we study the position of each actor within the interbank market network. The identification of individual contributions to the network activity is a crucial topic in the literature on social networks. Several measures have been proposed especially to detect the individuals important for the network, the so-called `key players' \citep[see][]{Borgatti:2006}. The available measures rely on different definitions of key players. For instance, centrality measures look at the structural importance of nodes \citep{Freeman:1979}. Social capital measures assess which individuals benefit most from a peculiar network structure \citep{Burt:1992, Borgatti:1998}. Key-player detection algorithms identify actors that contribute to cohesion and resource diffusion or to network disruption and fragmentation \citep{Borgatti:2006}.

In this paper, key players are defined as a multi-faceted concept. They are identified by a complex measure that combines different aspects of trading behavior. In modeling terms, each aspect corresponds to a network metrics. To capture all the information present in the data, we first examine these metrics separately. Then, we summarize all the information available into the key player measure. From an intuitive point of view, the idea is to measure the importance of a node by looking only at its direct ties, which indicate the lending (outgoing) or receiving (incoming) position of a node. For this purpose, we use the concept of degree centrality. This centrality measure identifies prominent nodes as those extensively involved in relationships with other actors. 

In detecting the key players, the first aspect to consider is the existence of actors that are particularly active on one or both sides of the market. In order to factor this into our analysis, we first consider a measure of dichomotous degree centrality. This measure counts the number of ties that are incident to a node or, equivalently, the number of nodes adjacent to it \citep[see][]{Freeman:1979}, given the information non-directly available from standard datasets. Tie weights are excluded. In directed networks, two measures of degree centrality can be computed, depending on whether incoming or outgoing ties are considered. These are the so-called `indegree' and `outdegree' centrality.

Indegree centrality looks at the number of actors that choose $i$ as a counterparty and lend money to it. This is a proper indicator of a bank's prestige or popularity. The reason is that it evaluates the reputation the market recognizes to $i$ as a trading partner. This is extremely important in the context of our analysis. Since our study focuses on a non-collateralized market, a measure of bank's reputation in the marketplace conveys relevant information. Indegree centrality is computed as: 
\begin{equation}\label{eq:app5}
k_{+i} = \sum_{j}{e_{ji}} 
\end{equation} 
where $k_{+i}$ ranges between 0, if $i$ has no incoming ties, and $(n-1)$. 

Outdegree centrality counts the number of nodes to which actor $i$ sends ties, and measures $i$'s trading activity. This statistics sheds light on a bank's capability to lend resources to other counterparties, and to establish and maintain  relationships. It takes the form
\begin{equation}\label{eq:app6}
k_{i+} = \sum_{j}{e_{ij}} 
\end{equation}
We compute the indegree and outdegree distributions for every year. Then, we examined their shapes, in order to verify the existence of fat tails, i.e. banks with a disproportionally high or low number of trading counterparties. In the literature on social networks, this analysis is usually performed by plotting the empirical distributions against the corresponding power-law ones \citep[see][]{Barabasi:1999}. The power-law distribution claims that the probability $P(k)$ that a node interacts with $k$ other nodes decays following $P(k)\propto k^{-\gamma}$ with an exponent $\gamma$ between 2.1 and 4. This shape of the probability distribution makes explicit the hypothesis that the networks are build over time through a preferential mechanism, with new actors exhibiting a higher probability to connect to more popular actors than to other actors. This feature leads to a `richer-get-richer' phenomenon, where highly-connected nodes (large $k$) have a large chance of occurring. Furthermore, this topological property tends to hold for large or complex networks \citep[e.g., see][]{walden}.   

The implications of a power-law shape for the degree distribution of ties for the interbank market are discussed in several studies of daily transaction, including  \citet{Iori:2008}. These authors observe that the presence of hubs makes the network structure extremely vulnerable to intentional attacks and epidemics. Attacks that simultaneously eliminate a small proportion of the hubs can propagate systemic risks and collapse a scale-free network. 

\begin{figure}[t]
\centering
 \caption{Indegree probability distribution with power-law fitting (2006-2009)}
  \label{fig:indegree}
 \subfigure[Indegree distribution for 2006\label{fig:gf2006}]
   {\includegraphics[width=0.4\textwidth]{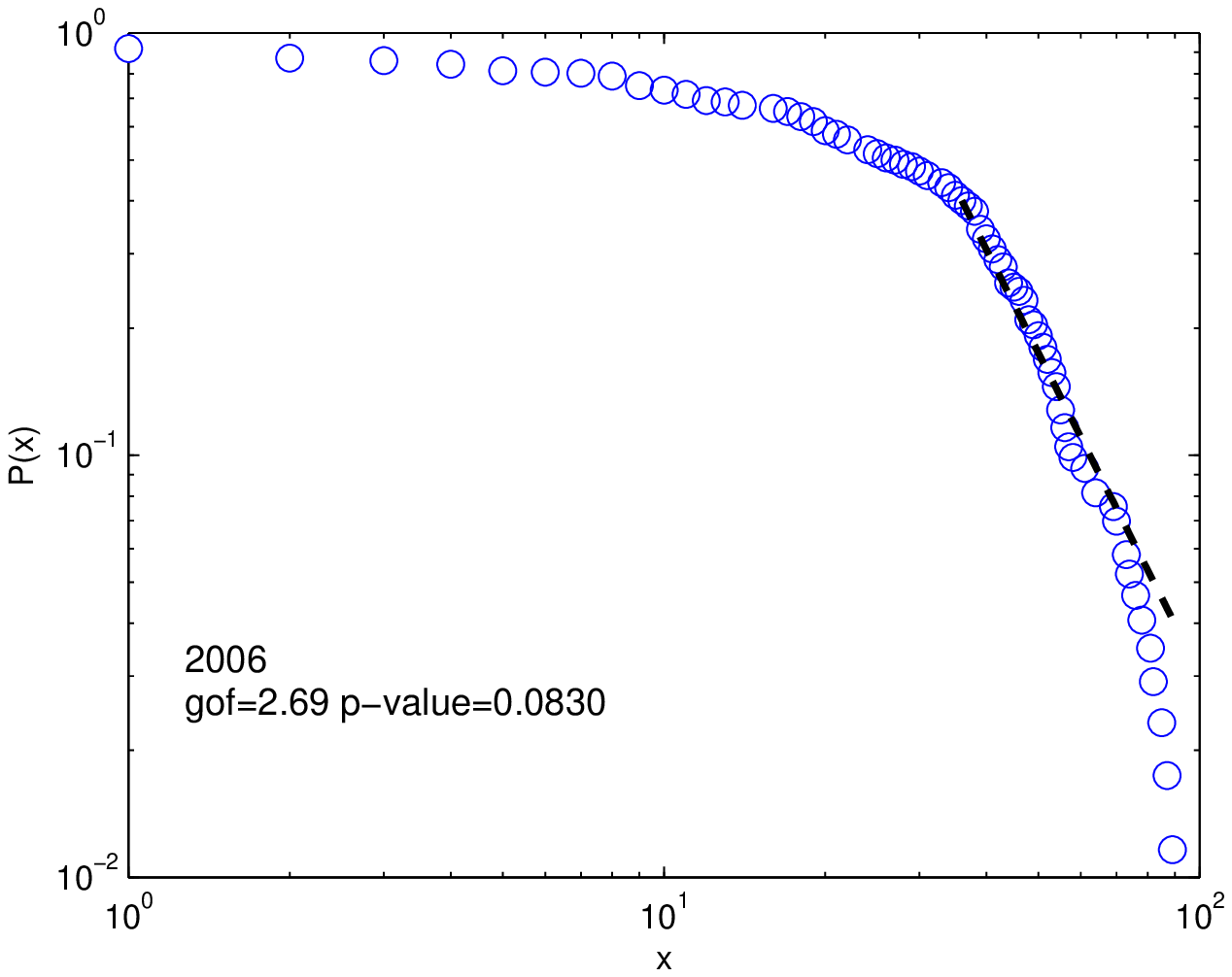}}\qquad 
 \subfigure[Indegree distribution for 2007\label{fig:gf2007}]
   {\includegraphics[width=0.4\textwidth]{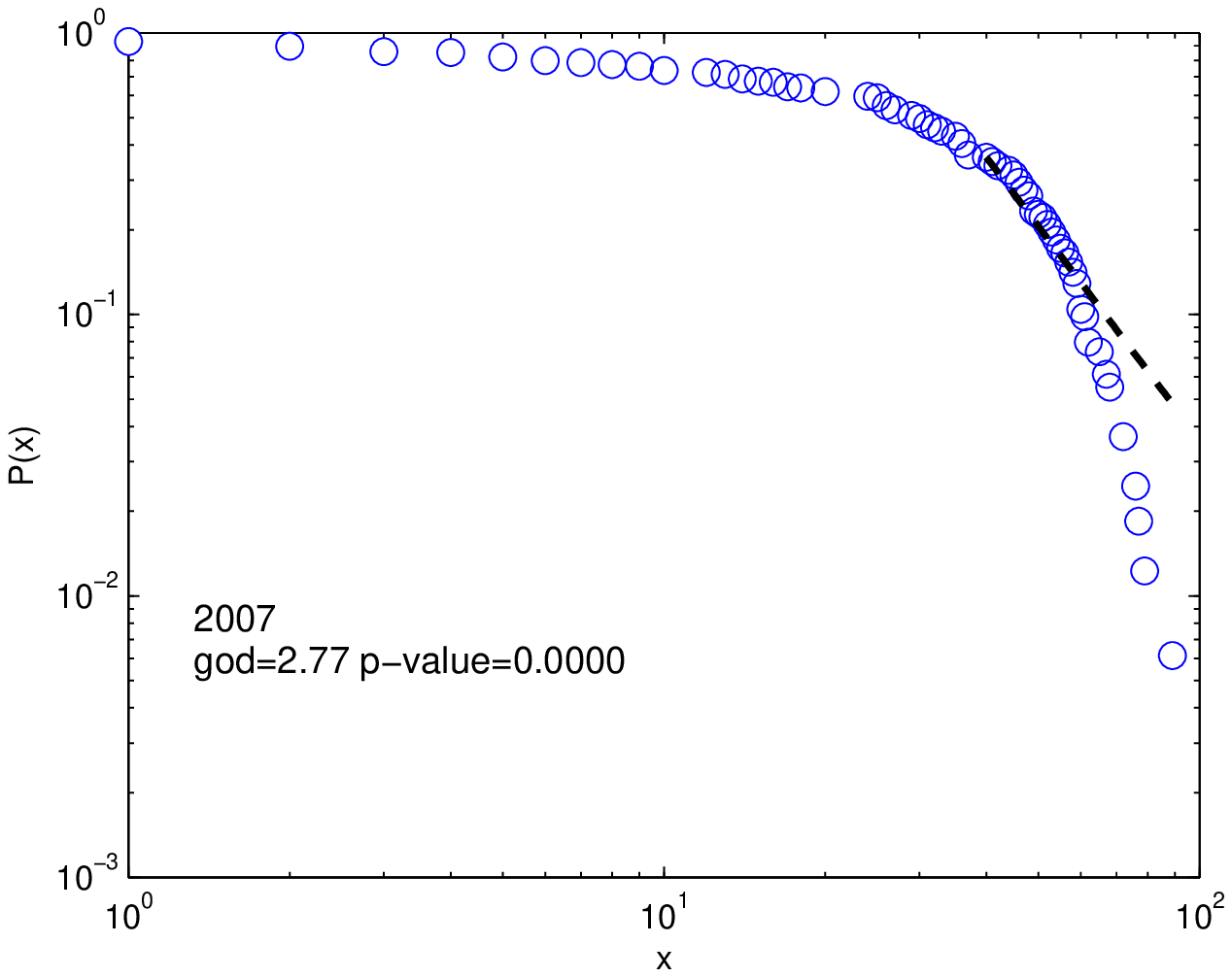}} \\ 
\subfigure[Indegree distribution for 2008\label{fig:gf2008}]
   {\includegraphics[width=0.4\textwidth]{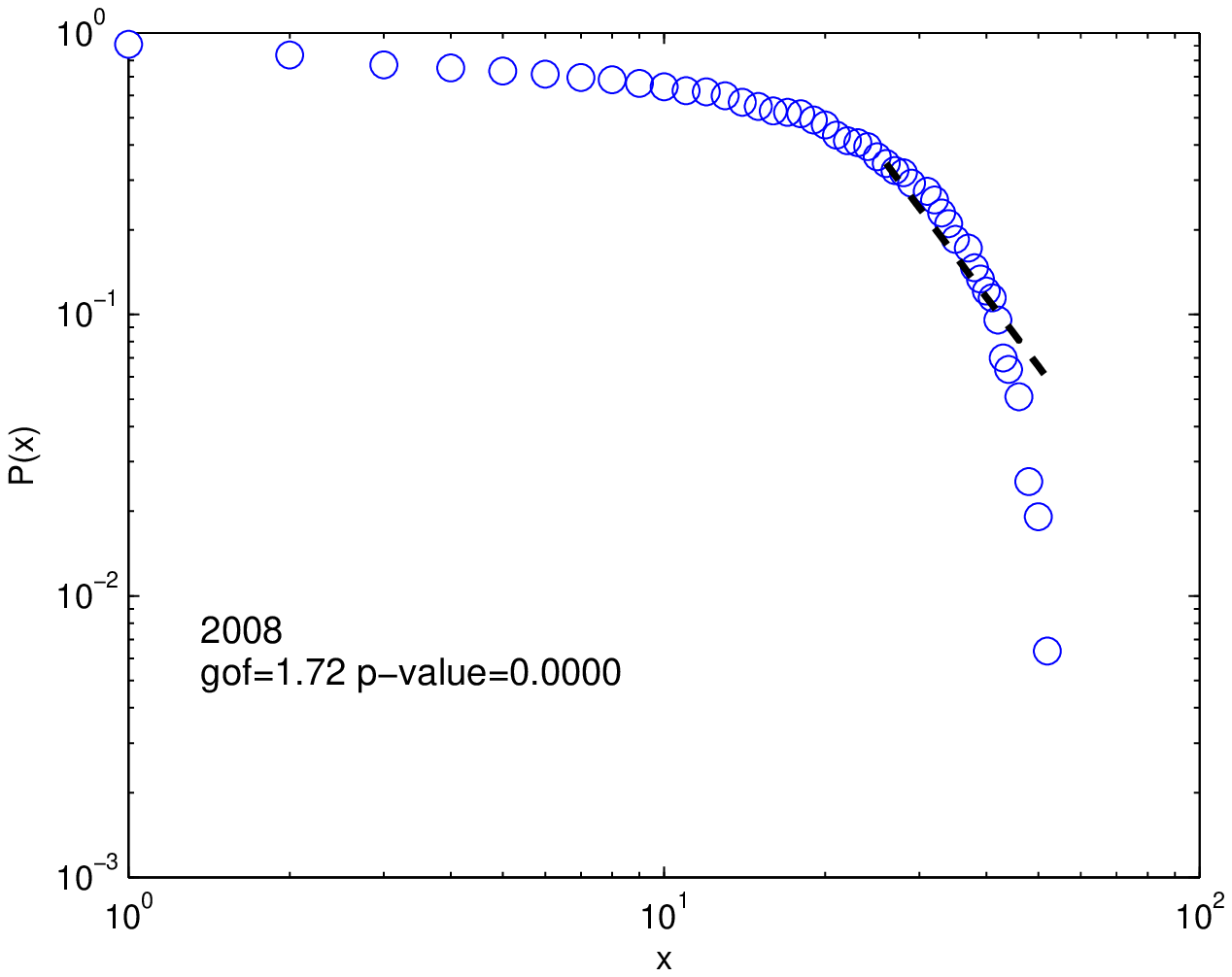}}\qquad 
 \subfigure[Indegree distribution for 2009\label{fig:gf2009}]
   {\includegraphics[width=0.4\textwidth]{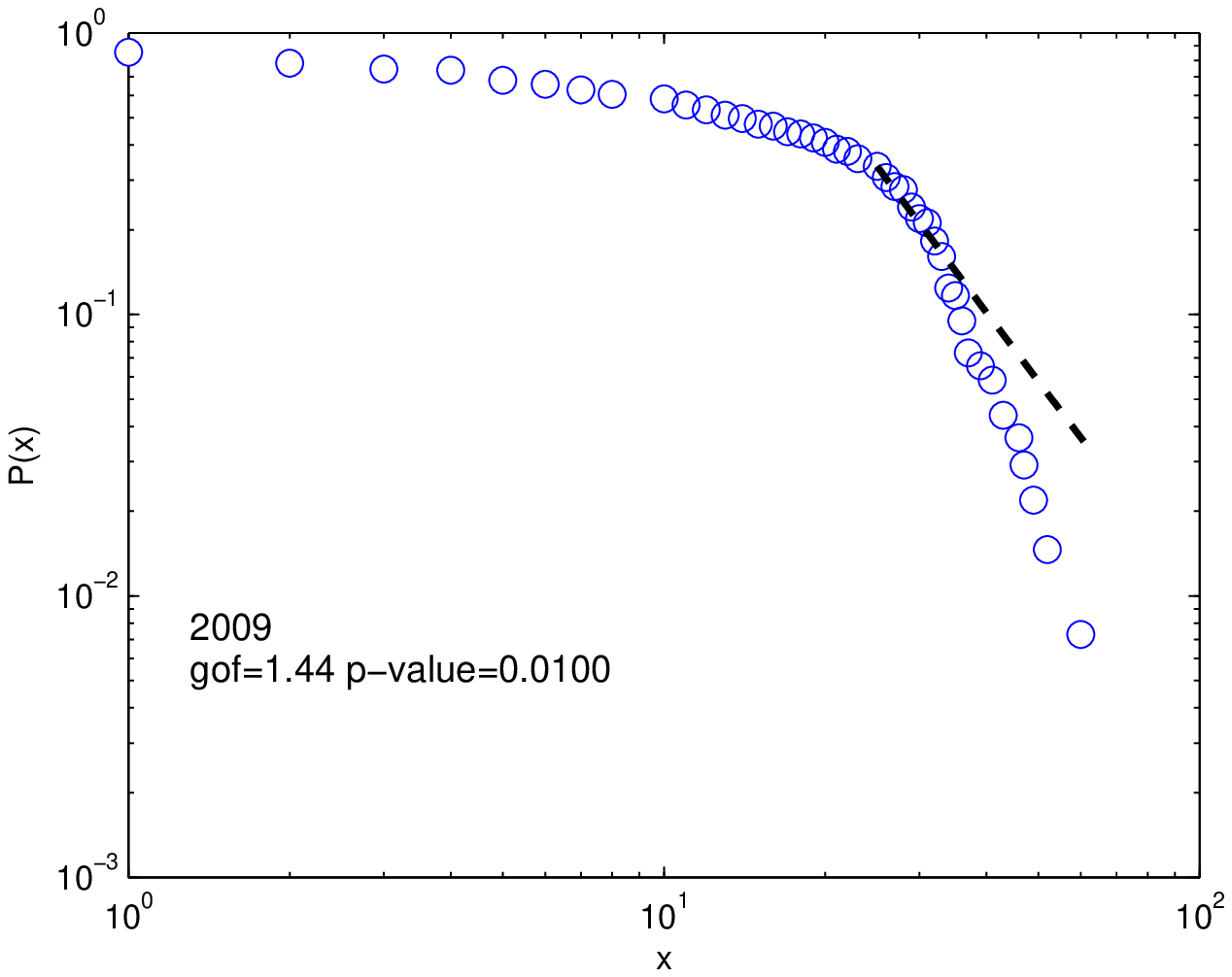}}

\changecaptionwidth{}\legend{Legend: On the horizontal axis there are the indegree (panel (a)-(d)) values and on the vertical axis the complement of the cumulative distribution function $P(X)=P(X\geq x)$. The points represent the observed values of the cumulative density functions and the dashed line the corresponding power-law. The goodness-of-fit between the data and the power law are calculated using the method described in \citet{Clauset:2009} and based on a Kolmogorov-Smirnov statistic. Since the resulting p-value is smaller than 0.1 the power law is not a plausible hypothesis for the data.}
  
\end{figure}

\begin{figure}[t]
\centering
 \caption{Outdegree probability distribution with power-law fitting (2006-2009)}
  \label{fig:outdegree}
\subfigure[Outdegree distribution for 2006\label{fig:gf20061}]
   {\includegraphics[width=0.4\textwidth]{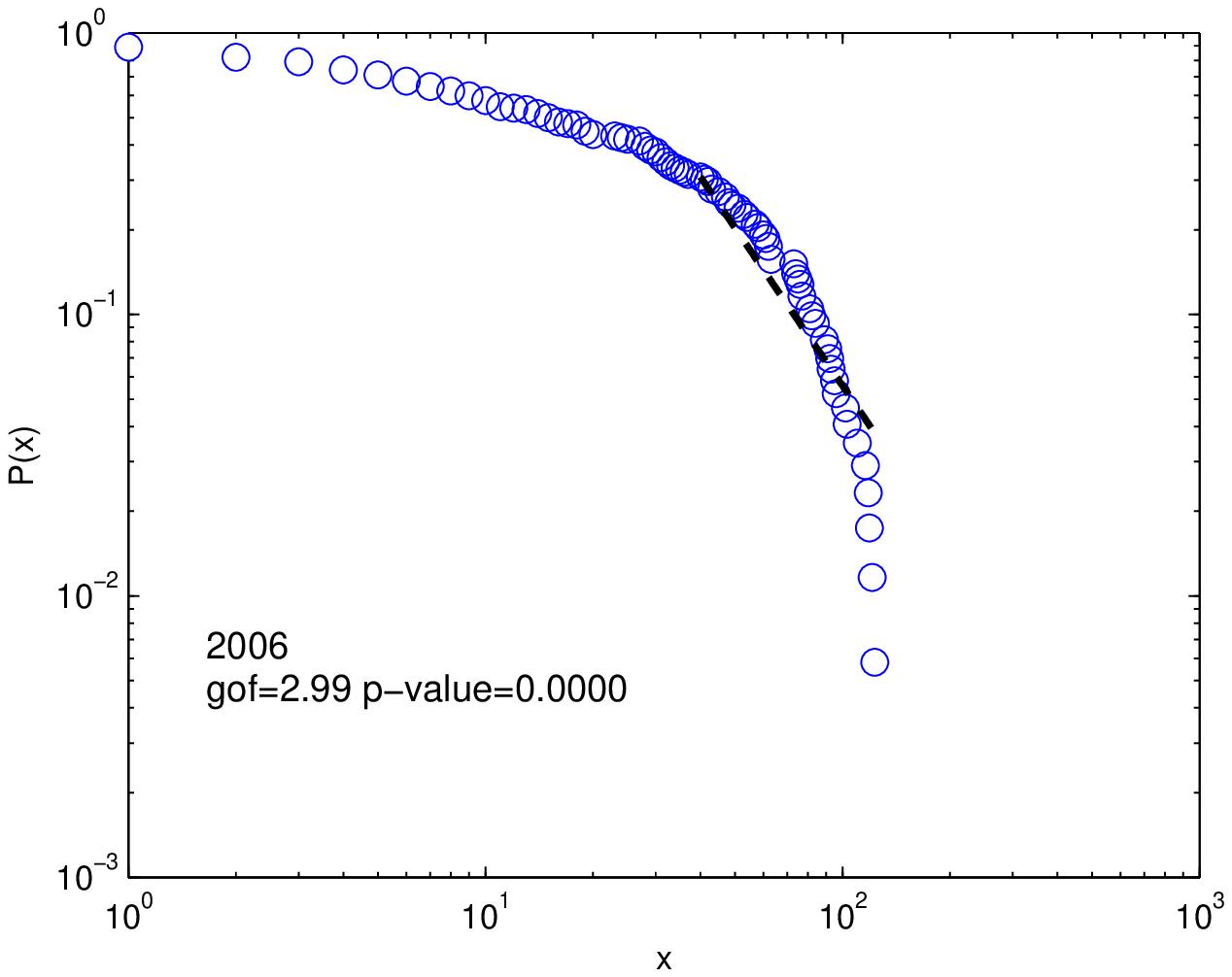}}\qquad 
 \subfigure[Outdegree distribution for 2007\label{fig:gf20071}]
   {\includegraphics[width=0.4\textwidth]{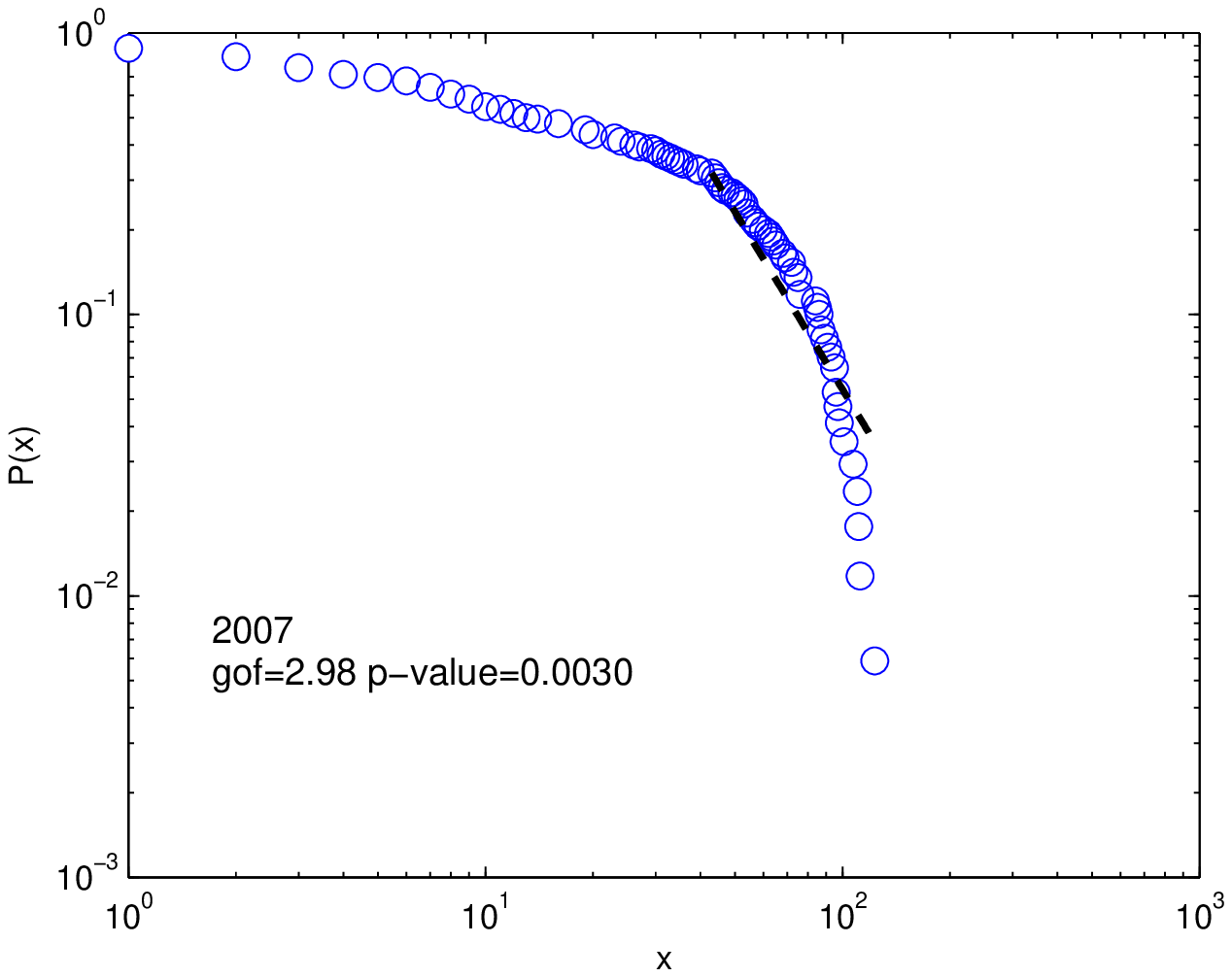}} \\
  \subfigure[Outdegree distribution for 2008\label{fig:gf20081}]
   {\includegraphics[width=0.4\textwidth]{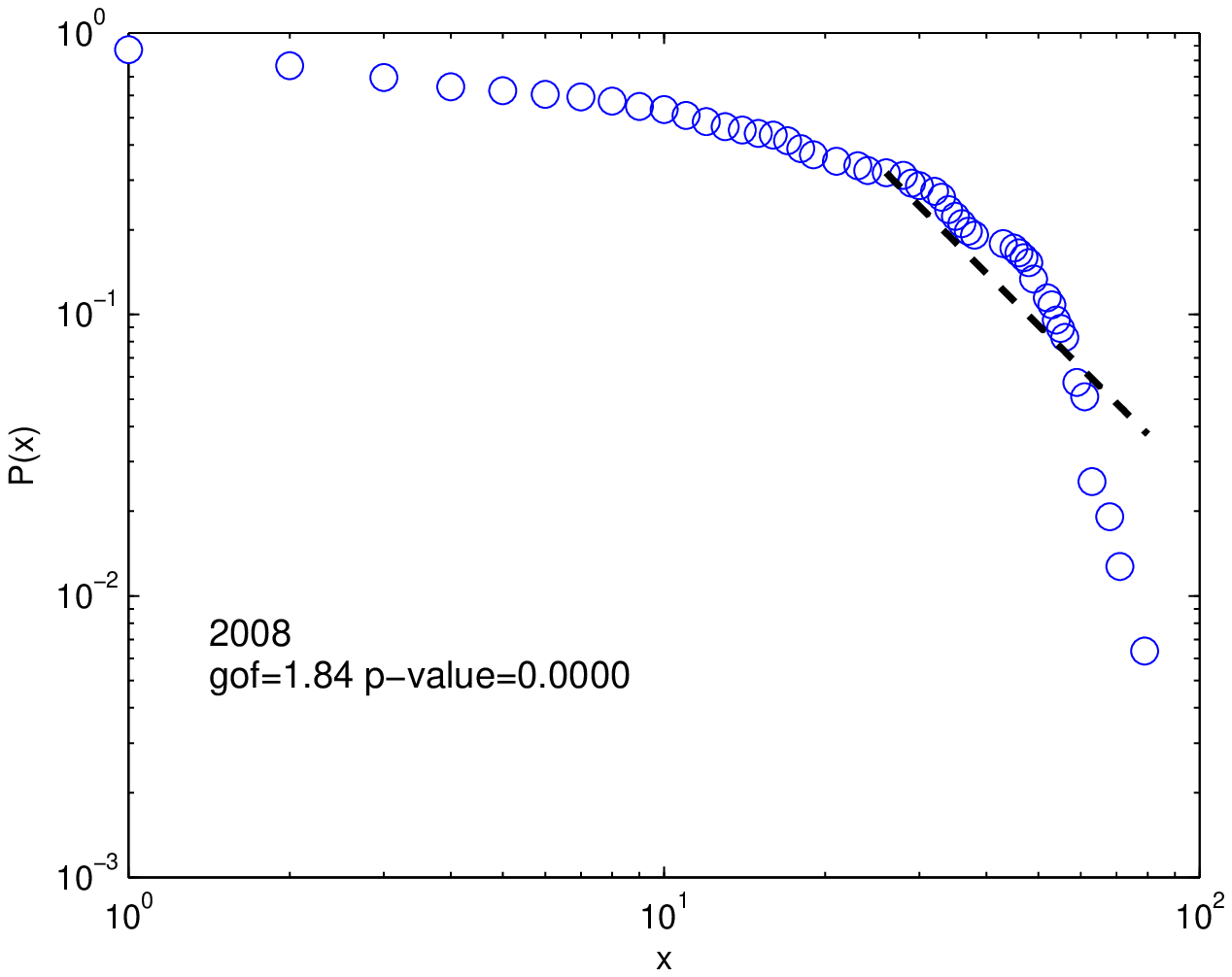}}\qquad 
 \subfigure[Outdegree distribution for 2009\label{fig:gf20091}]
   {\includegraphics[width=0.4\textwidth]{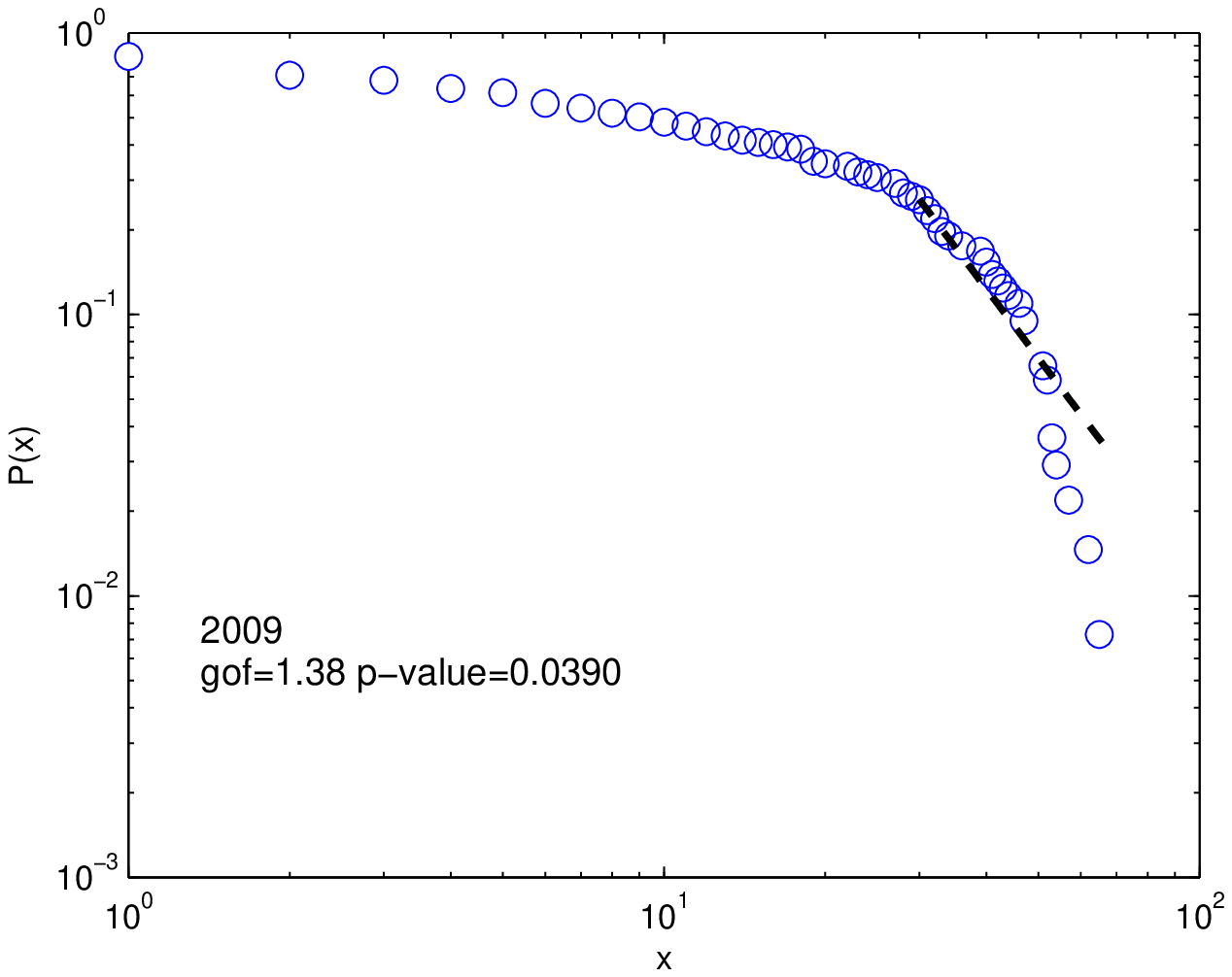}} 

\changecaptionwidth{}\legend{Legend: On the horizontal axis there are the outdegree (panel (a)--(d)) values and on the vertical axis the complement of the cumulative distribution function $P(X)=P(X\geq x)$.}
  
\end{figure}

Consistently with these studies, our results on yearly data show that both the distributions are heavily-tailed (see Figure \ref{fig:indegree} for indegree and Figure \ref{fig:outdegree} for outdegree distribution), although they do not follow a proper power-law. This shape of the distributions suggests the presence of a high degree of heterogeneity across banks in their trading behavior. Several banks exchange money with very few counterparties, whereas other institutions deal with many counterparties. The indegree distribution is especially right-tailed, suggesting that some banks borrow from at most 100 banks until 2008, and 89 in 2009. While the shapes of the in- and outdegree distributions remain almost the same over time, a deeper investigation of the descriptive statistics indicates that the absolute values have decreased fairly constantly, in line with the reduction of the market size. The average number of counterparties is 29.12 in 2006 and 29.27 in 2007, then falls to 25.08 in 2008 and to 20.24 in 2009.

Building on \citet{Bech:2008}, we then explore the financial strength of the actors. The question here is whether either the supply or the demand of deposits is controlled by a limited number of banks\footnote{The reader should notice that the high values of market concentration seem to point towards this.}. We study the distribution of the valued degree centrality, better defined as node strength \citep{Barrat:2004}. The strength of a node is the sum of $i$'s incoming or outgoing tie weights $w_{ij}$. It is an extension of node degree when analyzing weighted networks, and takes the form:
\begin{equation}\label{eq:app7}
s_{+i} = \sum_{j}{w_{ji}} 
\end{equation}
for incoming ties and 
\begin{equation}\label{eq:app8}
s_{i+} = \sum_{j}{w_{ij}} 
\end{equation}
for outgoing ties. Since we define the strength of a tie as the amount lent from the sender to the receiver over a given time-span, also the strength of a node sums up over these quantities. For brevity, the in- and outstrength distributions are not displayed here. However, we observe they resemble the related dichotomous degree version. They are both skewed and heavy-tailed, suggesting that few banks borrow or lend large amounts of cash. This is indeed the case especially for borrowing institutions, as demostrated by the preliminary analysis of Section 4.

\section{Our approach to the detection of the players\label{Sec_Dec} } 

From a methodological point of view, node strength and node degree are frequently complementary measures of node importance. For instance, the use of the weight distribution can lead to identifying nodes that, though having a small degree, mobilize and exert their control on a large amount of deposits. Thus, node strength need be not proportional to node degree. Hence, if we analyze the two measures separately, we may obtain incomplete information on node prominence. Like the correlation analysis of Table \ref{tab:cor} demonstrates, this is the case in our dataset to some extent. The Pearson's correlation coefficient between the node degree and strength for each year varies between 0.53 (in 2008) and 0.65 (in 2009) for outgoing ties, and between 0.62 (in 2009) and 0.75 (in 2007) for incoming ties.

In order to account for both node degree and strength, we compute the generalized degree centrality proposed by \citet{Opsahl:2010}. This index is the product between the number of nodes the actor $i$ is connected to and the average weight to these nodes. The generalised degree centrality is equal to:
\begin{equation}\label{eq:app9}
C^{w\alpha}_{D}(i) =k_{i}\times \left(\frac{s_{i}}{k_{i}}\right)^{\alpha}={k_{i}}^{(1-\alpha)}\times{s_{i}}^{\alpha}
\end{equation}
The constant $\alpha$ is a positive tuning parameter that determines the relative importance of the number of ties compared to tie weights. For $\alpha=0$, the value of the measure equals the degree centrality, whereas it equals the node strength for $\alpha=1$. Following \citet{Opsahl:2010}, we experiment with different values of $\alpha$. Overall, we find a substantial invariability of the node ranking to the specification chosen. We then adopt the most conservative approach and set $\alpha=0.5$. In doing so, we assign the same positive importance to node degree and strength in identifying key players. The high values of Pearson's correlation coefficient between the distributions of the generalized degree centrality with $\alpha=0.5$, the degree centrality on one side, and the node weight on the other for both outgoing and incoming ties (see Table \ref{tab:cor}) suggest the appropriateness of $C^{w\alpha}_{D}(i)$ as an indicator of node importance within the network.

Our definition of key players represents a synthesis of the outgoing (lending) and incoming (borrowing) ties. We assume (and attempt to verify) that the actors of the interbank liquidity market play different roles, i.e. some mainly lend while others mainly borrow. Some actors in particular, the key players, control the resource flow, by lending (or borrowing) large amounts of money to (from) many counterparties. Then, they are large liquidity providers or borrowers. The former role takes the form of a combination of lending large volumes to many banks, and of borrowing small volumes from few others. The latter role takes the opposite combination. Hence, from a computational point of view, the key players are identified by a high absolute value of the difference between the generalized out- and in-degree centrality ($C^{w\alpha}_{O-I}(i)$). If this difference is positive, the key player is a liquidity provider. Otherwise, it is a liquidity borrower. The presence of actors with particularly high values of $C^{w\alpha}_{O-I}(i)$ would represent the definite proof of market distortions in the distribution of liquidity. Therefore, this would also demonstrate that the interbank market is not a perfect market.

\begin{figure}[!p]
 \centering
 \caption{Distributions of net liquidity providers (2006-2009)}
 \label{fig:oi}
 \subfigure[2006\label{fig:oi1}]
   {\includegraphics[width=0.42\textwidth]{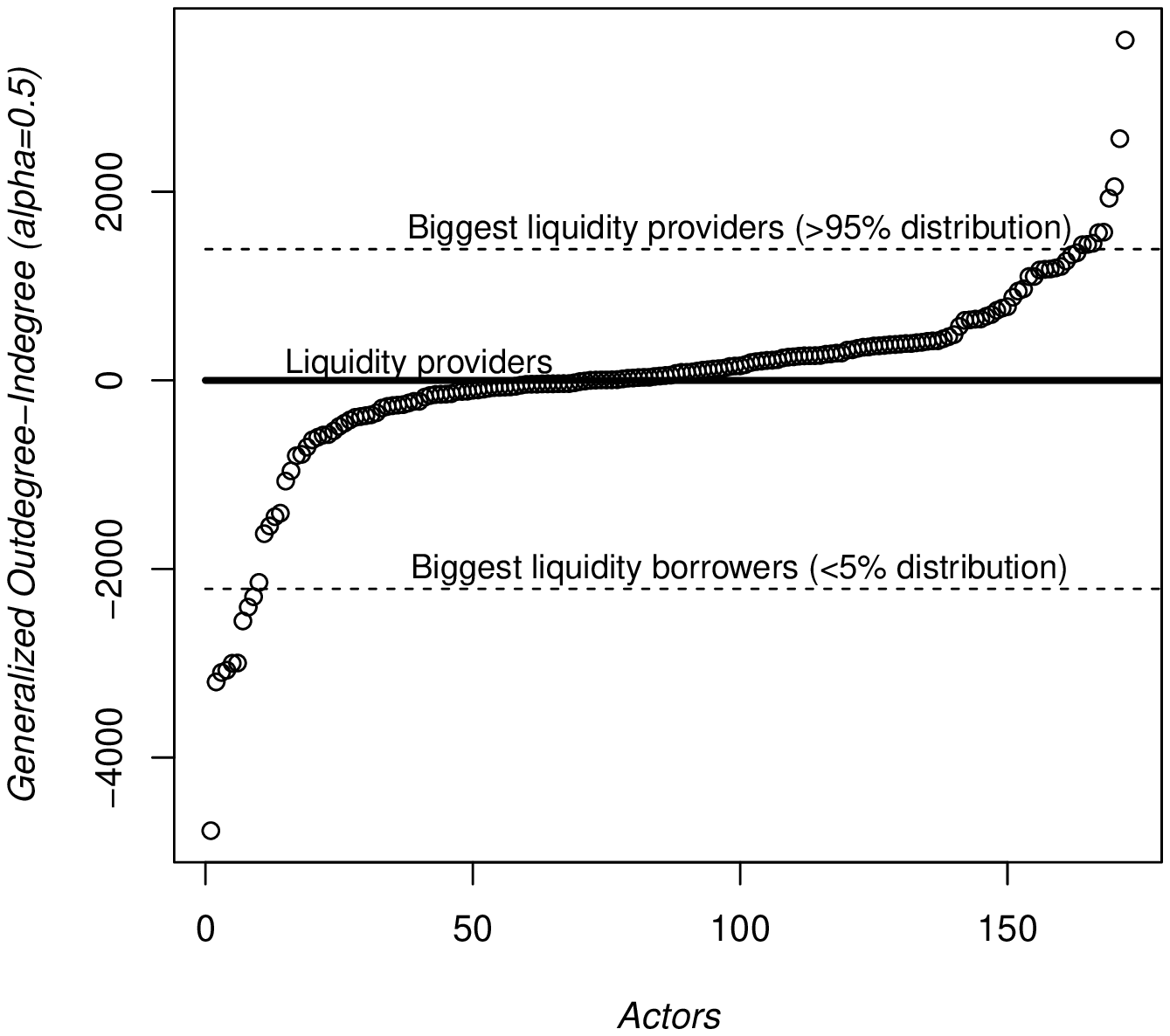}}
 \subfigure[2007\label{fig:oi2}]
   {\includegraphics[width=0.42\textwidth]{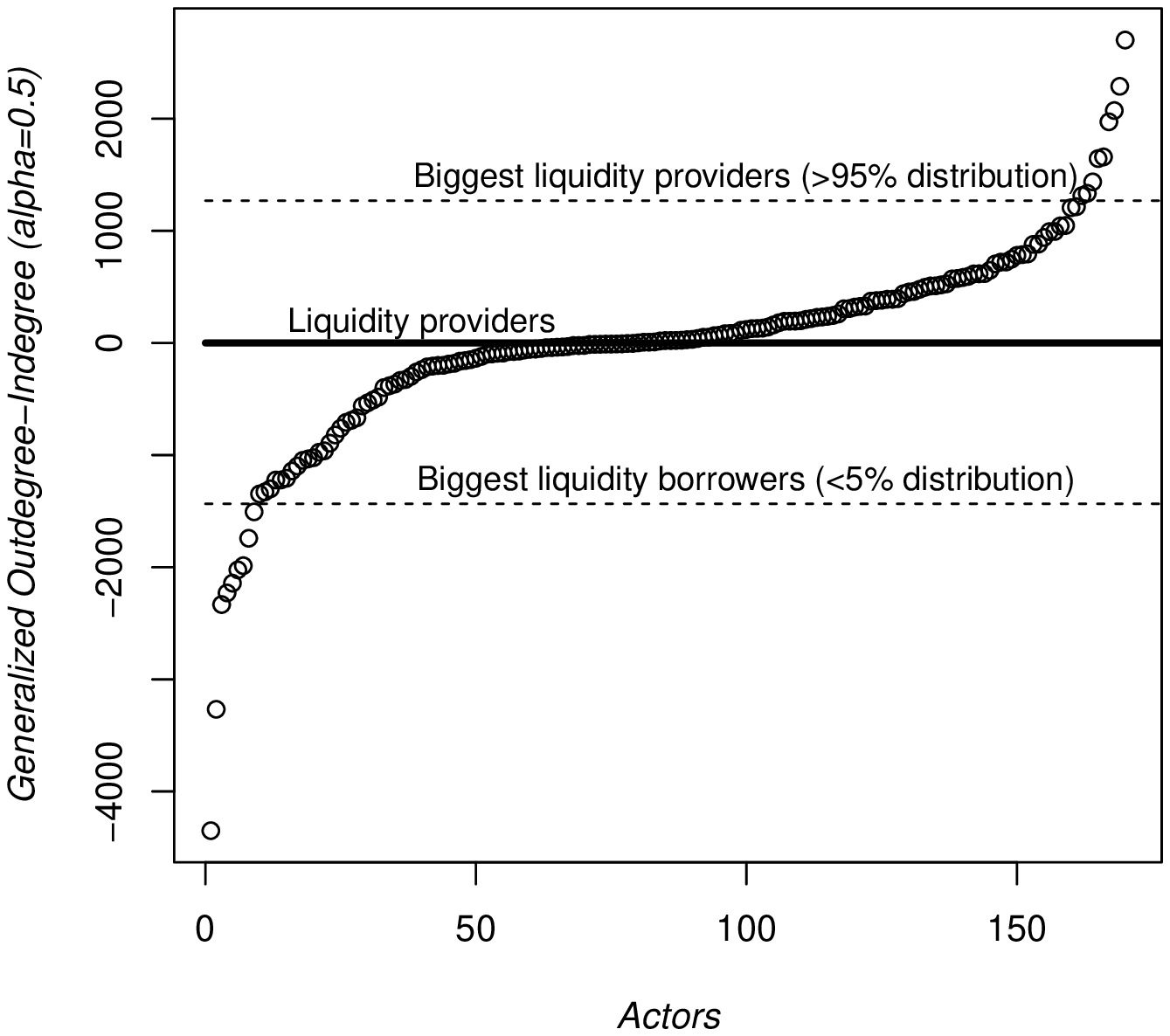}}
   \\
 \subfigure[2008\label{fig:oi3}]
   {\includegraphics[width=0.42\textwidth]{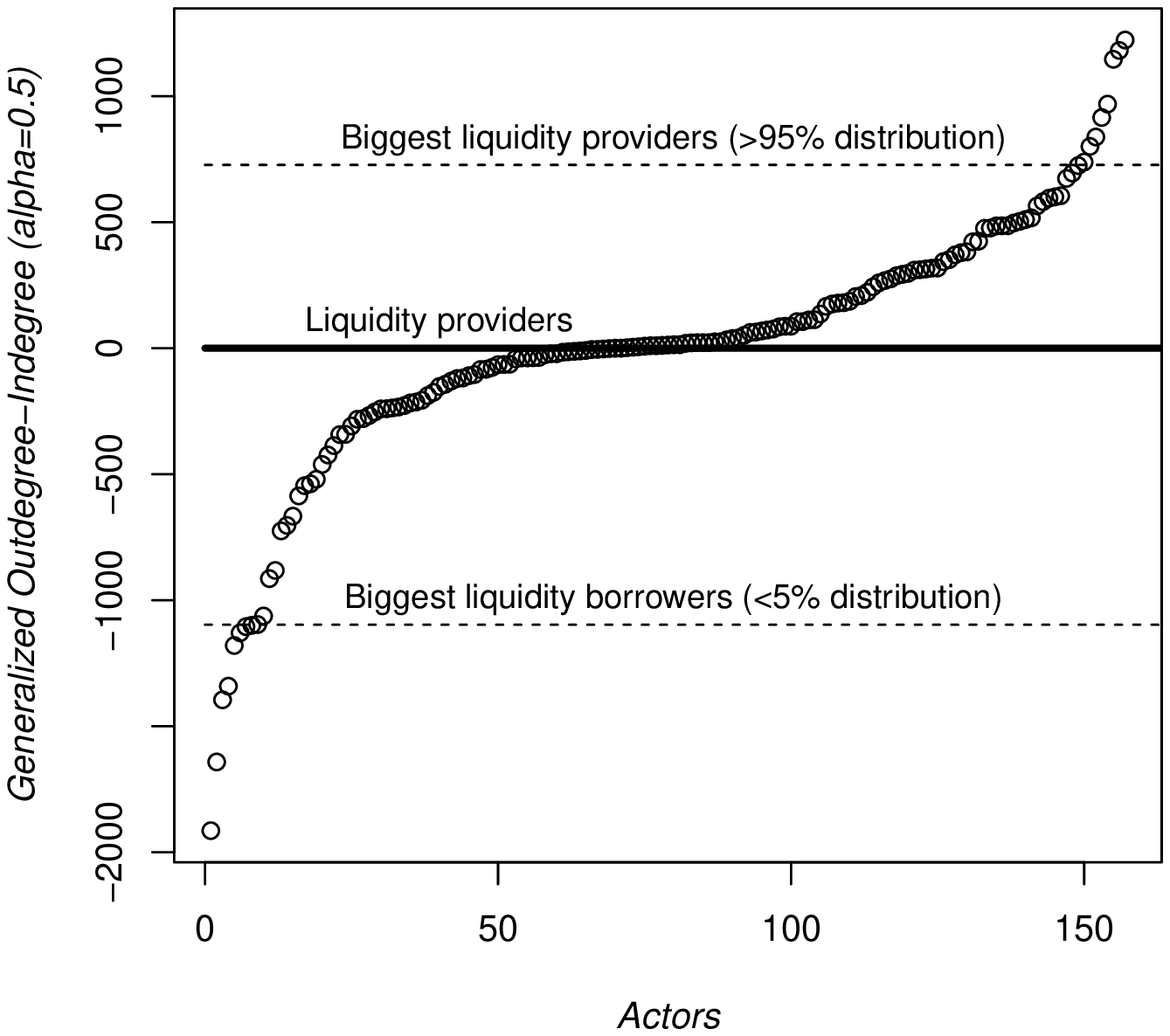}}
 \subfigure[2009\label{fig:oi4}]
   {\includegraphics[width=0.42\textwidth]{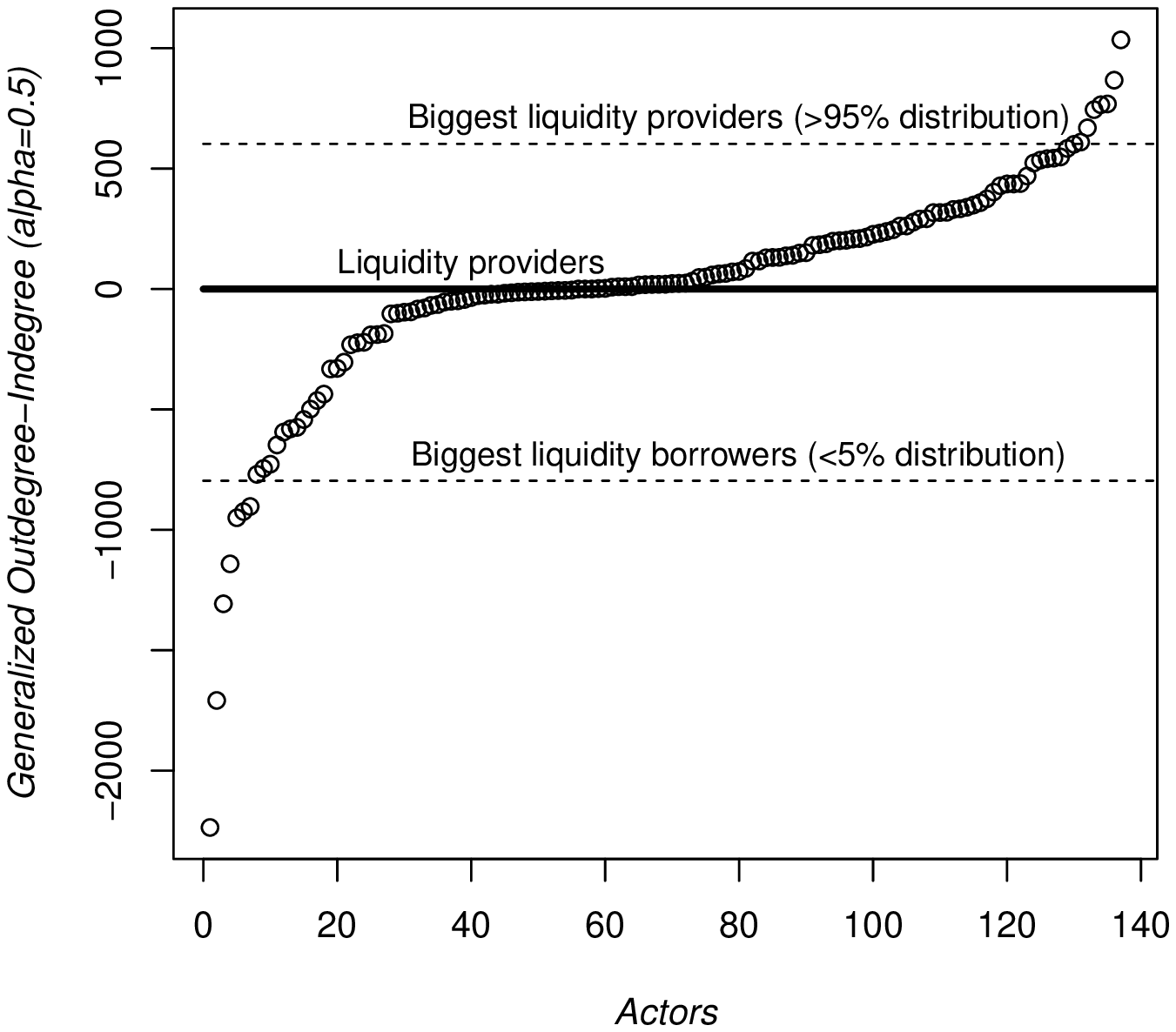}} 
\changecaptionwidth{}\legend{Legend:The graphs have the actors on the horizontal axis and $C^{w\alpha}_{O-I}(i)$ on the vertical axis.}
\end{figure}

Figure \ref{fig:oi} displays the $C^{w\alpha}_{O-I}(i)$ distribution for the observation period. For all the years, the distribution exhibits a similar shape, which is fairly symmetric around 0, i.e. the neutral position. The first part of the distribution is flat, then increases steadily, following a linear function, and finally very sharply in the tails. Around 35\%-40\% of the actors falls in a very small interval around 0. Therefore, they do not have a definite role nor market power, but lend and borrow around the same amount of money and to/from a similar number of counterparties. Their market share is less than 0.2\% in absolute value. If we look at the steadily increasing part of the distribution, we find banks which display a prevailing behavior as lenders or borrowers, although fairly weak. Over each observed year, they account for the 35\%-40\% of the market and their market share is smaller than 1.0\%. Then, the most part of liquidity is traded by a very small number of banks and both sides of the market are characterized by the presence of banks with a clear role. This confutes the assumption of lack of distortions in the liquidity distribution of the interbank market. The increase in the values of $C^{w\alpha}_{O-I}(i)$ and also in the market shares is particularly sharp after the $95th$ percentile (and before the $5th$ percentile). As a consequence, we set a threshold $t_{1}$=$95th$ percentile and define as large liquidity providers the banks with $C^{w\alpha}_{O-I}(i)\geq t_{1}$. Then, we set $t_{2}$=$5th$ percentile and consider as large liquidity borrowers the institutions with a value of $C^{w\alpha}_{O-I}(i)\leq t_{2}$. According to our definition, these banks are both key players.

\section{The market role of `big providers' and `big losers'}

Identifying which nodes play a key role in a network is a difficult task. This is especially the case in financial markets where the opportunistic behavior of market players is difficult to detect.  The descriptive measures proposed in the previous section point to the presence of inefficiencies and asymmetries in the interbank market. However, the opacity of the information available does not allow to provide a structural interpretation of these patterns. 

In this section, we provide a detailed discussion of the characteristics of the key players. We start by focussing on the 95th percentile of the distribution \ref{eq:app9} . This provides information on the `big providers', namely on the banks that offer a large portion of the liquidity in the market while borrowing little. We then discuss the evidence on the 5th percentile of the distribution \ref{eq:app9}. This tail identifies the `big losers', which are the key drivers for the net demand for liquidity

The empirical distributions reported in Figure \ref{fig:oi} are characterized by an invariant right tail for 2006 and 2007. A similar picture emerges between 2008 and 2009. The nodes change, though, between 2007 and 2008, thus indicating that an important change takes place in 2007. We can interpret this finding as evidence that the eruption of the financial market turmoil the Euro area in August 2007 changed the composition of the market\footnote{Tensions in segments of the US Dollar-denominated money markets reached their highest point on August 9 2007. In order to stabilize the market conditions, the ECB started a series of open-market operations supplying Euro-denominated liquidity.}. 

Who are the key players and how do they behave from an individual point of view? To shed light on the structure of exchanges, we report some descriptive statistics on the lending activities of the big providers in Table \ref{lending_big_providers_to_market}, and of the big losers in Table \ref{lending_big_losers_to_market}.\footnote{To avoid approximation errors, we report statistics from the original tick-by-tick dataset in this section.} There are several questions of interest. The first one concerns the market share covered by the big players. In other words, we should understand how truly important the big players are in the supply of liquidity. The descriptive statistics from Table \ref{lending_big_providers_to_market} show that these actors can indeed be dubbed as `big'. For instance, in 2006 and 2007, the percentage of liquidity they cover exceeds 33\% of the total amount present in the market. This suggests that the key players have a tight grip on traded volumes both on the lending side and the demand side. Table \ref{lending_big_losers_to_market} provides evidence on the supply-side coverage of the big losers. By construction, the largest net borrowers lend only up to 5\% of the market. 

What about the drivers for the demand side of e-Mid? Tables \ref{borrowing_big_providers} and \ref{borrowing_big_losers} report some descriptive statistics on the borrowing activities of the big providers and the big losers, respectively. The market is characterized by some big providers that never borrow. In 2008, the largest net lenders borrow for less than 1\% of the market. On the other hand, the big losers control no less than 40\% of the borrowing side.

How stable is the composition of the group of key players? Tables \ref{lending_big_providers_to_market} and \ref{borrowing_big_losers} reveal a rich landscape of behavior. First of all, the identification of the key players confirms the marked changes in the structure of the networks that has taken place during the turmoil period. There is substantial dynamics of entry and exit in the groups of big providers and big losers key across time. For example, only 67\% of the losers are the same from 2006 to 2007. This figure drops to 44\% in 2008. This confirms our previous results indicating that the interbank market is not composed by actors with long-term strategies. Rather, large net-lending and net-borrowing decisions appear contingent on temporary factors, and are driven only by opportunistic behavior and short-term strategies\footnote{Differently from the big providers, the big losers have only Italian counterparties.}. 

The changing ranks at the top layer of the lending relations may be due to several factors. As stressed by \citet{holthausen:1}, liquidity hoarding takes place for precautionary motives during phases of market breakdown. Hence, banks that are large net liquidity suppliers at a given point in time may choose to reverse their course of action and turn into net borrowers. The banks may have even chosen to stay out of the market to avoid the adverse consequences of the increase in system-wide counterparty risk. The entry-exit dynamics from the group of big losers is, instead, largely affected by an issue of stigma in large demands for cash \citep[see][]{vento}. Since posting ask trades in e-Mid generates information available to the whole market, banks have the incentive to leave the platform during phases of market turbulance, and opt for over-the-counter trading.

\subsection{The pricing behavior of the key players}

The results from the previous section indicate that liquidity is not distributed evenly across participants in the interbank market. Both the supply and the demand side of e-Mid are controlled by a small number of banks. This is indeed a distortion in the organization of the market. In this section, we investigate whether these structural distortions are also a source for trading frictions. We study the pricing implications arising from the division of roles between net lenders and net borrowers. 

\begin{figure}[!t]
\begin{center}
\caption{Borrowing rates paid by the big losers on contracts with four major categories of counterparties (2006-2009)}
\label{r_borrowing}

\subfigure[Borrowing rates for 2006\label{fig:br2006}]
   {\includegraphics[width=0.5\textwidth]{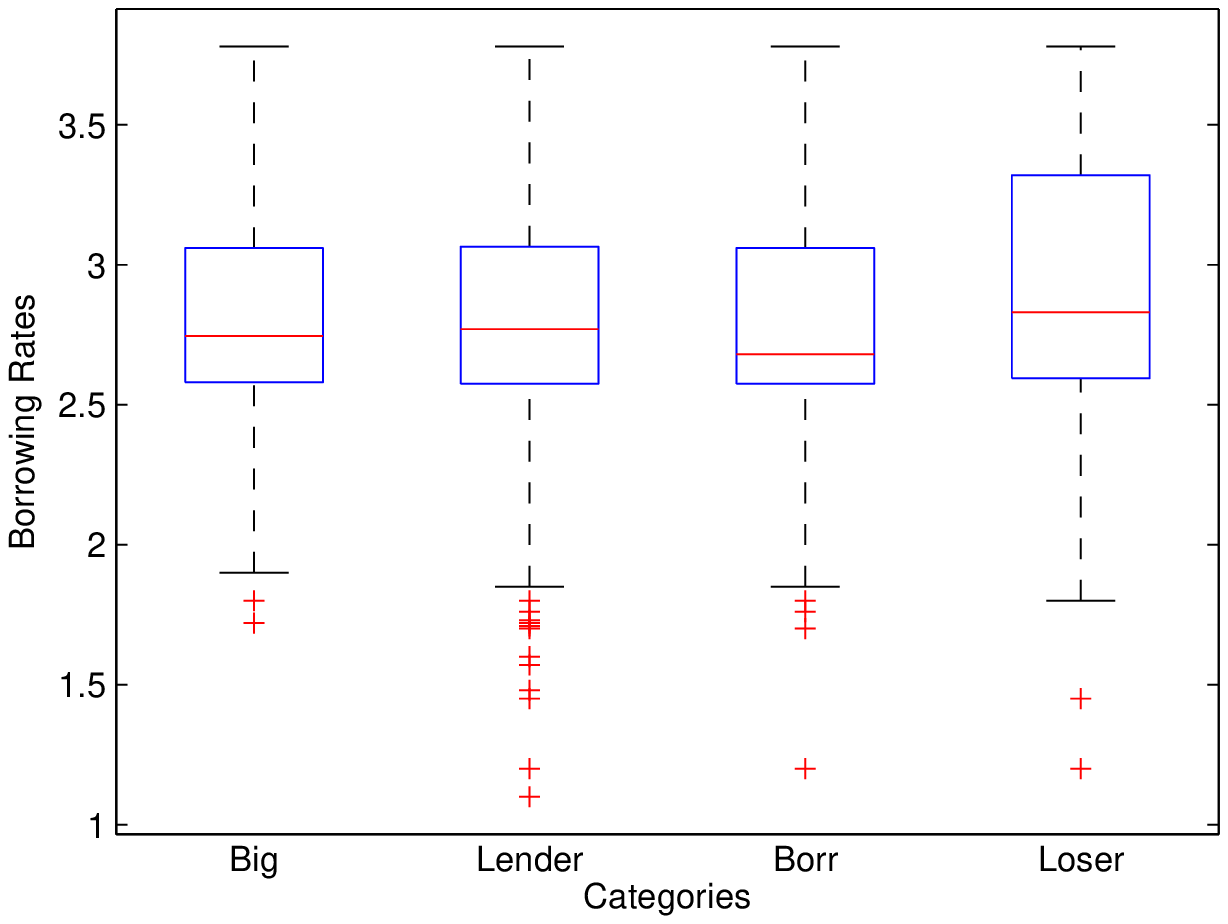}}%
 \subfigure[Borrowing rates for 2007\label{fig:br2007}]
   {\includegraphics[width=0.5\textwidth]{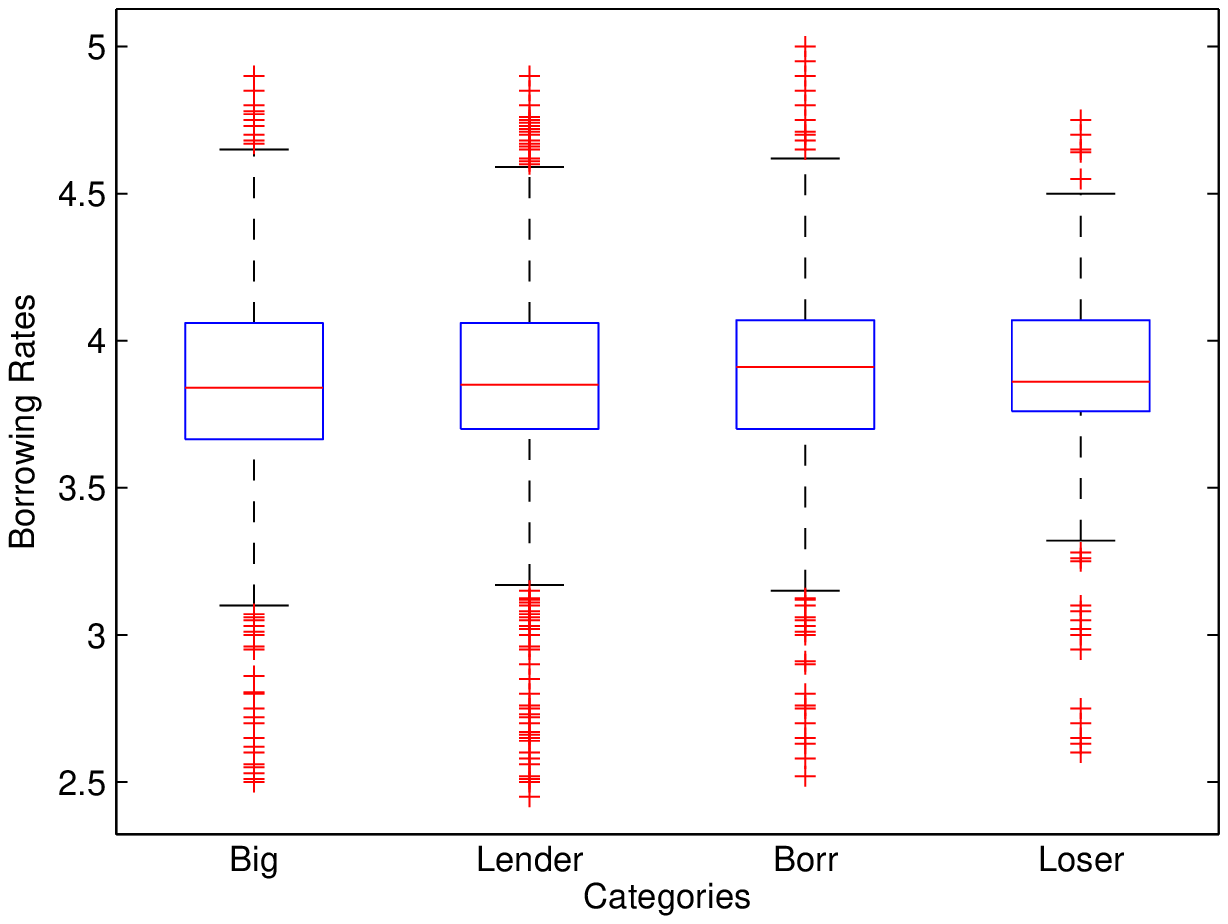}}\\
  \subfigure[Borrowing rates for 2008\label{fig:br2008}]
   {\includegraphics[width=0.5\textwidth]{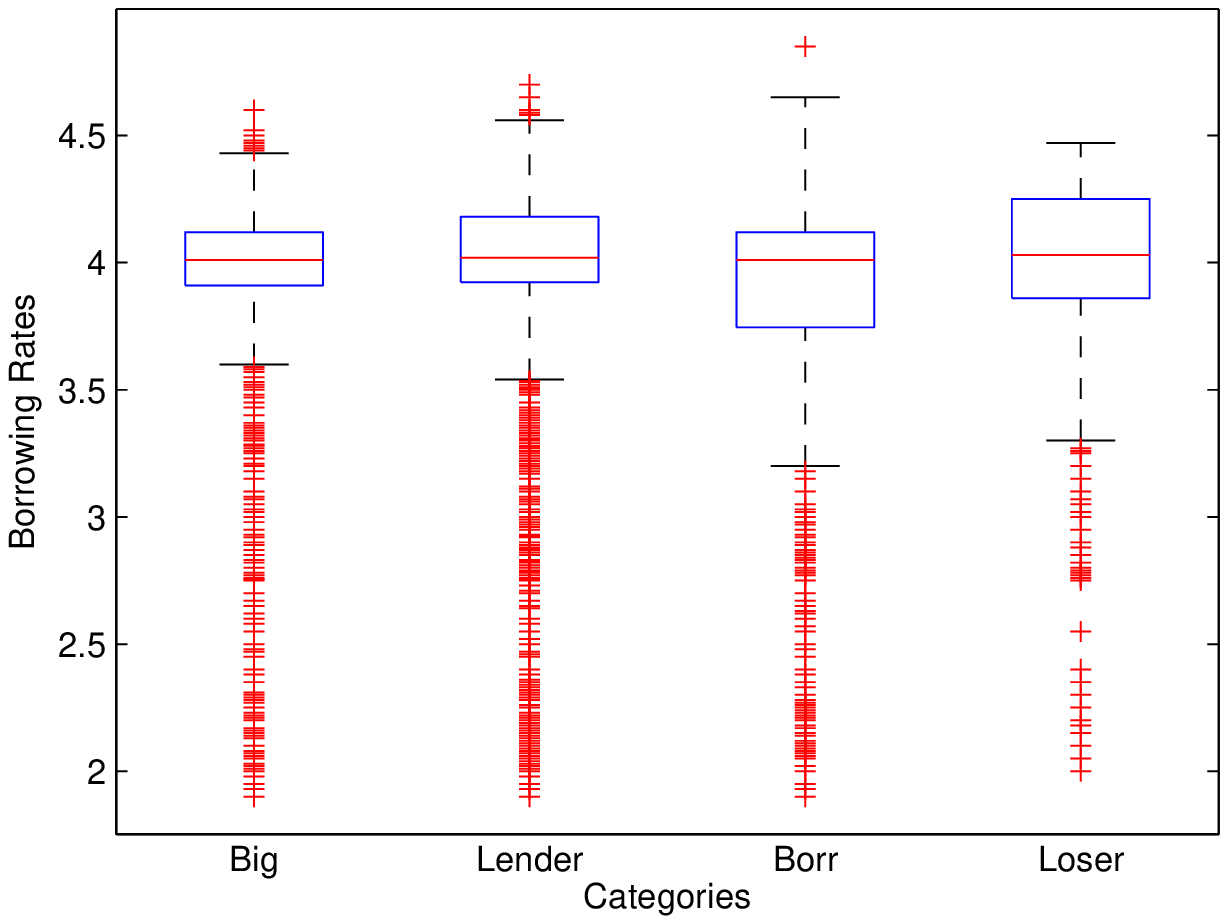}}%
 \subfigure[Borrowing rates for 2009\label{fig:br2009}]
   {\includegraphics[width=0.5\textwidth]{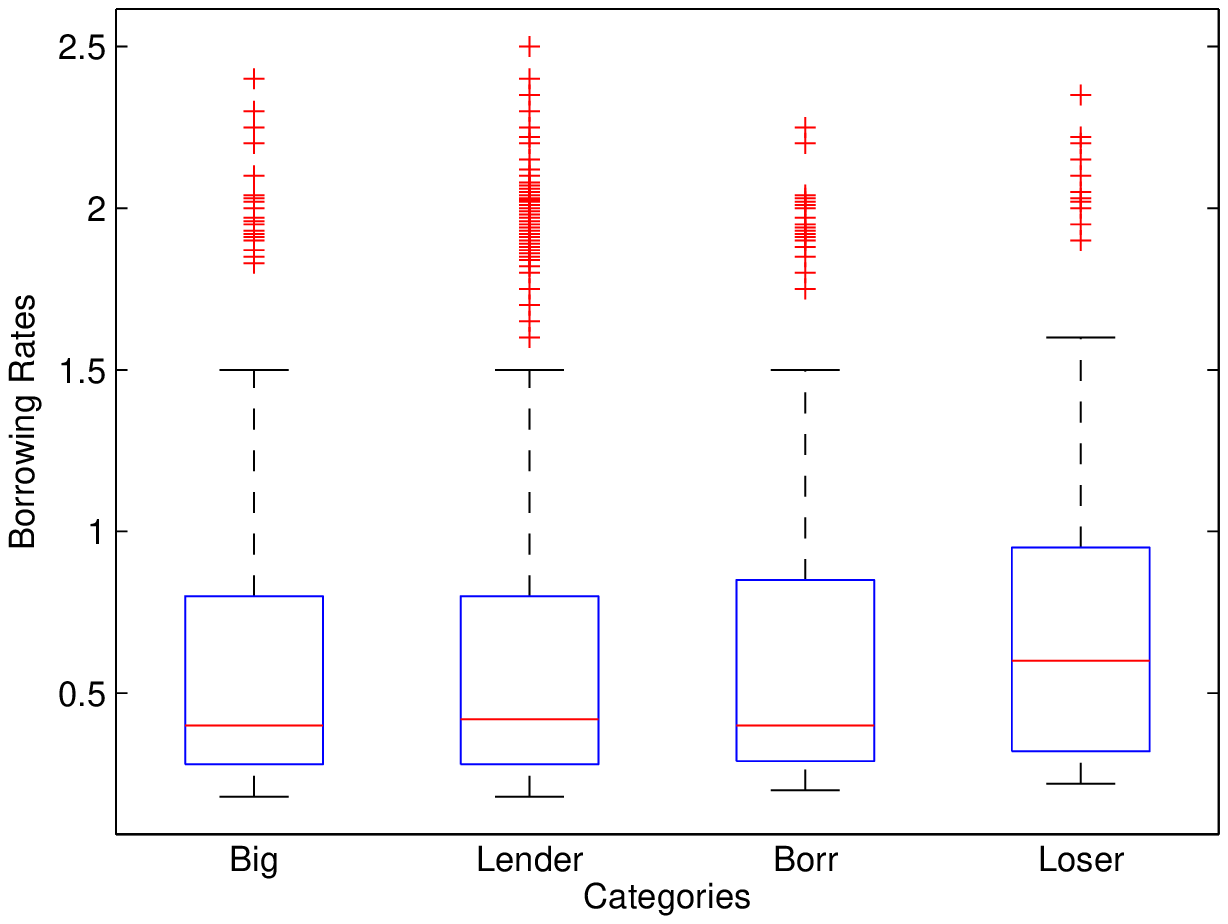}} 

\changecaptionwidth{}\legend{Legend: The term `Big' denotes the big providers. `Loser' indicates the big losers. `Lender' and `Borr' refer, respectively, to net lenders and net borrowers that do not fall on the tails of distribution \ref{eq:app9}.
Categories of counterparties are derived by means of $C^{w\alpha}_{O-I}(i)$ distribution, setting target thresholds as follows: for 'Big'  $C^{w\alpha}_{O-I}(i)\geq t_{1}$; 'Lender' $0 \leq C^{w\alpha}_{O-I}(i)< t_{1}$; 'Borr' $t_{2} < C^{w\alpha}_{O-I}(i)< 0$; 'Loser'  $C^{w\alpha}_{O-I}(i)\leq t_{2}$.}
\end{center}
\end{figure}

\begin{figure}[!t]
\begin{center}

\caption{Lending rates offered by the big providers on contracts with four major categories of counterparties (2006-2009)}
  \label{r_lending}

\subfigure[Lending rates for 2006\label{fig:ld2006}]
   {\includegraphics[width=0.5\textwidth]{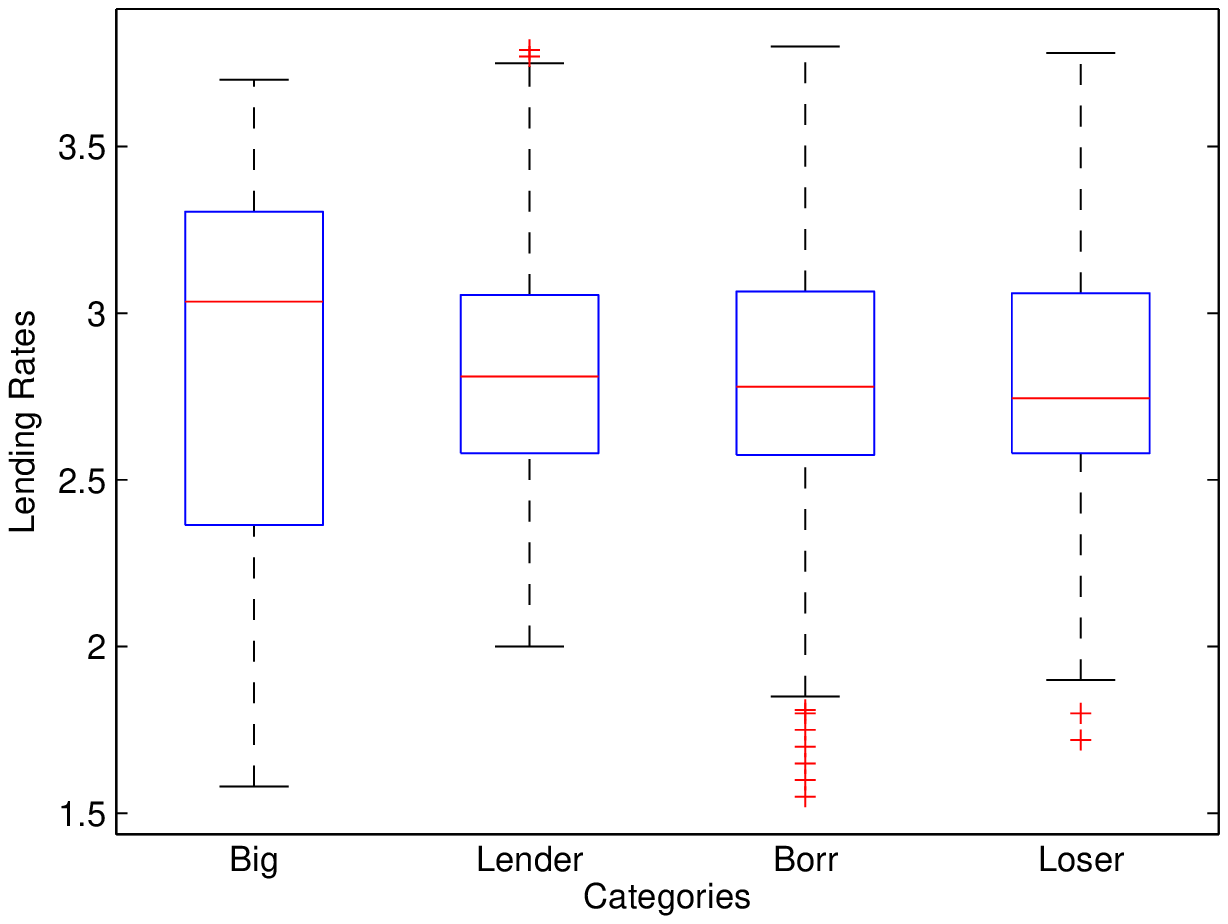}}%
 \subfigure[Lending rates for 2007\label{fig:ld2007}]
   {\includegraphics[width=0.5\textwidth]{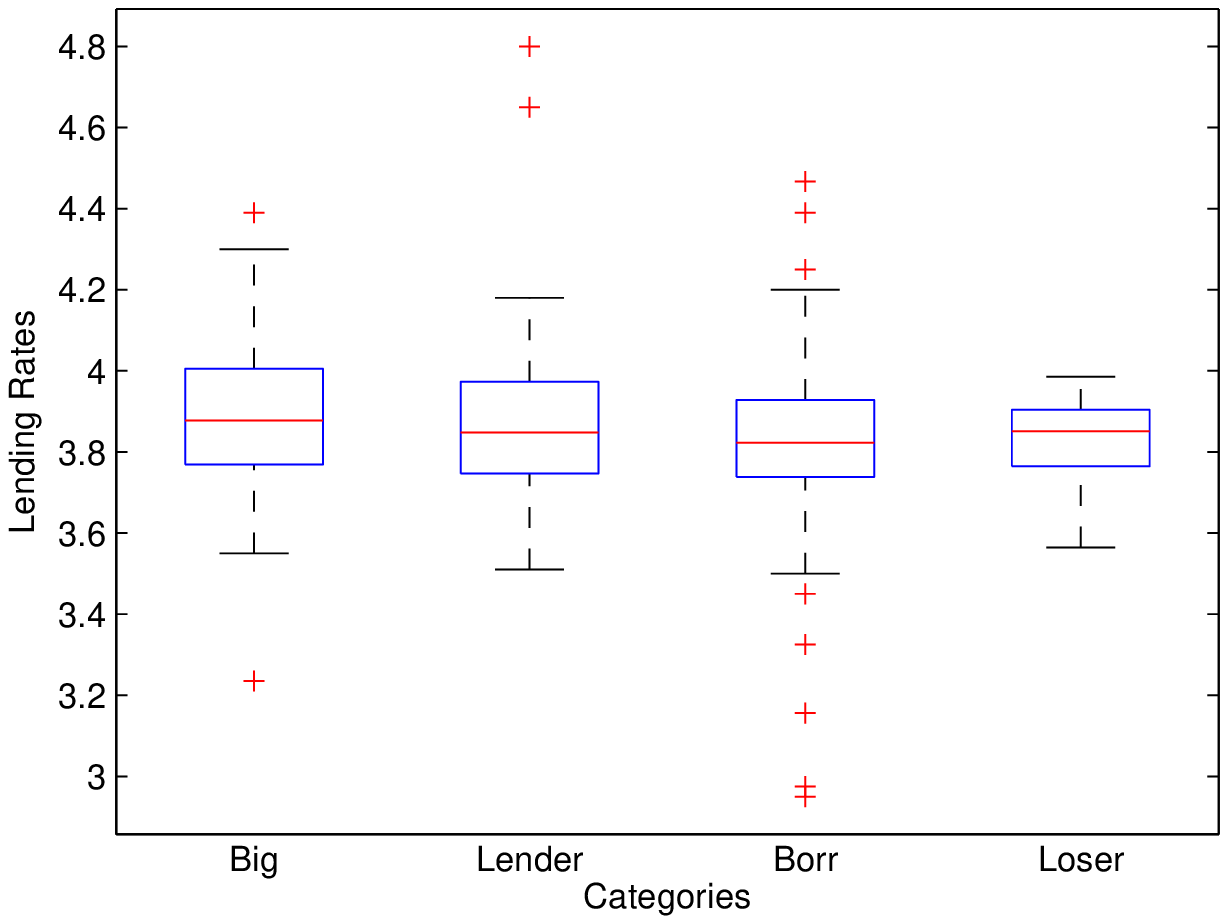}}\\
  \subfigure[Lending rates for 2008\label{fig:ld2008}]
   {\includegraphics[width=0.5\textwidth]{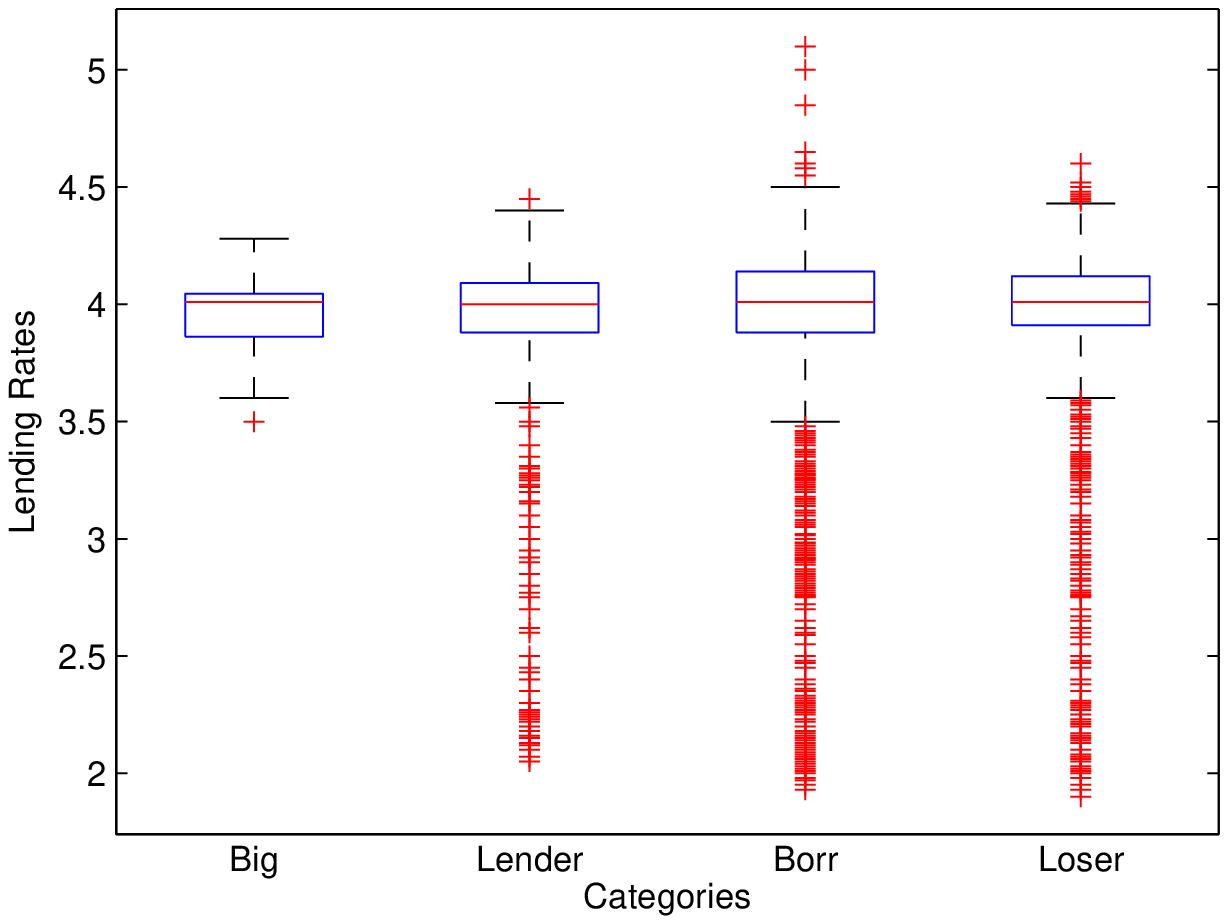}}%
 \subfigure[Lending rates for 2009\label{fig:ld2009}]
   {\includegraphics[width=0.5\textwidth]{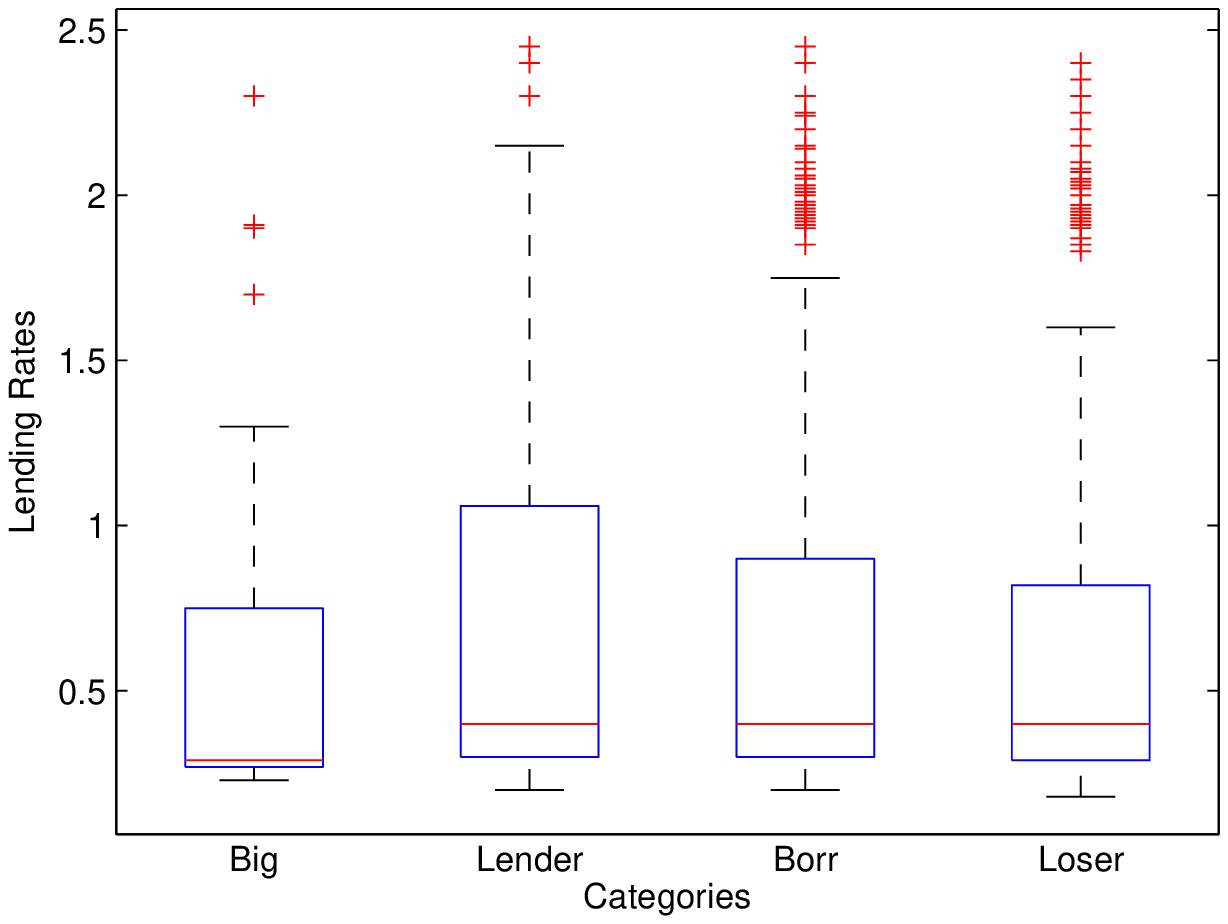}} 

\changecaptionwidth{}\legend{Legend: The term `Big' denotes the big providers. `Loser' indicates the big losers. `Lender' and `Borr' refer, respectively, to net lenders and net borrowers that do not fall on the tails of distribution \ref{eq:app9}.
Categories of counterparties are derived by means of $C^{w\alpha}_{O-I}(i)$ distribution, setting target thresholds as follows: for 'Big'  $C^{w\alpha}_{O-I}(i)\geq t_{1}$; 'Lender' $0 \leq C^{w\alpha}_{O-I}(i)< t_{1}$; 'Borr' $t_{2} < C^{w\alpha}_{O-I}(i)< 0$; 'Loser'  $C^{w\alpha}_{O-I}(i)\leq t_{2}$.}
\end{center}  
\end{figure}

Our first question of interest is related to the cross section of borrowing costs. By how far do the costs for funds vary across banks? At what interest rates do the big providers lend, in comparison with the borrowing rates of the big losers? Figure \ref{r_borrowing} reports a box plot of borrowing rates paid by the big losers. We report information on contracts with four different types of counterparties. We consider the big providers and the big losers, on two opposite sides. We also consider as relevant counterparties two groups of net lenders and borrowers that do not fall either on the 95\% or the 5\%-tail of the distribution \ref{eq:app9}. Figure \ref{r_lending} plots similar statistics for the lending rates offered by the big providers\footnote{We should stress that Figures \ref{r_borrowing} and \ref{r_lending} report data from tick-by-tick transactions, rather than yearly average numbers.}.

The median borrowing rates are rather similar for the different categories of banks between 2006 and 2008. This pattern changes in 2009 though, when the median rates for the big losers are markedly higher than for the other institutions. This may that the worsening of conditions in the interbank market after the Lehman Brothers bankruptcy has generated an uneven impact on banks borrowing conditions. In particular, in the distribution of borrowing rates for big losers, there is a majority of banks that pay high interest rates in 2009. However, statistical tests for the difference of means of the borrowing rates provide results that are not significant at standard confidence levels. In other words, we find no statistical backing for the hypothesis that the big losers face borrowing rates in 2009 that are higher than for other sample periods\footnote{We perform the hypothesis tests on the mean rates with a $p$-value level of 0.05. We have also computed test statistics for the difference of means across categories of banks for each year. The null hypothesis of no difference is, again, accepted in all the cases. The acceptance of the null is caused by the large variability of the interest rates for each category of bank.}. 

Figure \ref{r_lending} suggests that also the variation of the median lending rates across types of net lenders or borrowers is rather limited over the sample period. The difference between the median lending rate demanded by the big providers and by the other lenders is highest in 2006, before the beginning of the turmoil. Also in this case, tests for the difference of means between categories of lenders in each year does not reveal any statistically significant pattern. 

The big providers enjoy market power over the distribution of liquidity. In other words, their behaviour determines how the money supply of the ECB propagates through e-Mid. Hence, we would like to understand if the big providers exploit their market power by imposing pricing policies that are more aggressive -- or `predatory' -- than those characterizing the other lenders by charging high lending rates. Two alternative and `extreme' hypothesis may be proposed. On one hand, a bank that controls the relative supply of deposits within a network enjoys the market power to impose lending rates higher than the average rate. In this case, we could think of a network as a monopolistically-competitive market where profit-maximizing banks generate profits in excess of a market with perfect competition. On the other hand, exchanges within a network may take place because of trust among counterparties. A lender may even avoid charging above-market rates to secure a `safer' demand for funds that carries a low counterparty risk. For instance, this may happen when a lender enjoys a persistent excess of cash holdings that raise the internal cost of capital. 

The hypothesis that a big provider is an aggressive lender for the big losers is not corroborated by the data. The difference between the interest rates charged to be big losers and the average market rates are positive and statistically significant only for 33\% of the big providers in 2006. This share does not exceed 50\% in 2007 and 2008. Owing to the overall dry-up of the market throughout the turmoil period, this fraction drops to 29\% in 2009. 

Strikingly the big losers behave as predatory lenders when they supply liquidity to the market (see Table \ref{lending_big_losers_to_market}). In 2006, 5 big losers out of 9 lend at an average rate that is markedly higher than the market rate. The time variation of the lending rates is somewhat complex though. Overall, we find evidence of large shifts in the lending rates charged by the big losers across time. The patterns are characterized by a rich dynamics. For 2006 we observe a positive correlation between borrowing rates and the number of network nodes that is equal to 0.881. In other words, the lending rates increase as a function of the number of possible counterparties within the reach of network relations. To put it differently, a wider liquidity supply does not contribute to lower the lending rates demanded by the big losers in 2006. The pricing policies change in 2007, when the probability to find a counterparty does not affect the probability of demanding high lending rates. In fact, for this period, the correlation between the borrowing rates and the frequency of exchanges within a network is equal to -0.011. At the same time, the differences of pricing strategies are strongly correlated with the intensity of exchange relations\footnote{In this case, the correlation between borrowing rates and number of trades is large and equal to 0.62.}. Despite a sizeable increase in lending rates in 2008, the frequency of network exchanges becomes negatively correlated with the interest rates\footnote{The correlations between the borrowing rates, on one hand, and the number and frequency of trades on the other are -0.45 and -0.39, respectively.}. This pattern is completely reversed in 2009 when, again, an aggressive pricing policy is highly correlated with the chance to exchange within a network of larger size.

\subsection{A discussion of the economic implications of our results}

The identification of key players that control the demand and supply in the interbank market stresses the role of distortions in the distribution of money supply through the banking system. We find that e-Mid is not a frictionless marketplace as it is not characterized by perfect competition among banks. Moreover, we document a disconnect between the demand-supply imbalance of traded volumes and the market determination of prices. One would expect the forces that drive the interaction between demand and supply of funds to affect the prevailing level of the interest rates too. However, over the long run, that does not take place in e-Mid. In the pricing of interbank loans, the relative size of the counterparties in terms of transacted volumes does not matter. Rather, both the lending and the borrowing rates are determined by factors that are not necessarily related to centrality of a bank in the market. 

There are two obvious considerations in this domain, which are both related to the nature of prices. Interest rates in the interbank market are tied to the monetary policy stance of the ECB as set, for instance, in the interest rates on the main refinancing operations (MRO). In addition, counterparty risk is a key component that contributes to the market spread from the MRO rates \citep[e.g., see][]{holthausen:1}.

The implications of these findings for the formation and evaluation of liquidity supply by the ECB is compelling. Banks need liquidity to carry out their daily operations. In presence of consolidated network links between financial institutions, a buoyant supply of liquidity by a central bank may not necessarily `pass through' the system due to the presence of the key players. Interbank rates are, however, largely affected by conditions unrelated to traded volumes. In other words, even though market conditions as measured by the lending rates may appear to ease, the patterns of traded volumes need not display a substantial improvement in periods of market distress. 

This discussion suggests that evaluating the impact of the expansionary liquidity policy carried out by the central banks during the recent financial crisis should take into account the persistent distortions characterizing the market already before August 2007. Under this view, the structural characteristics of e-Mid discussed in this paper may cast doubt on the effectiveness of measure for liquidity easing implemented by central banks.

\section{Conclusion}

Available studies on network effects in the money market study short-term relations between banks. In this paper, we focus on the long-term patterns of network formation. Using a dataset from the electronic platform e-Mid, we provide evidence of evolving relations that induce structural distortions in the market during the recent market turmoil. These patterns of exchange are largely asymmetric and imply a clear division of roles between lenders and borrowers. We identify key players that affect the demand and supply sides of the market, and consider the implications of their pricing strategies. Strikingly, we find that only a fraction of the key players use their market power to impose aggressive pricing policies on counterparties during the turmoil period. This suggests that structural distortions do not necessarily translate into trading frictions. 

Our results provide relevant economic insights for the conduct of the liquidity policy of the ECB. The asset exchanged in the interbank market has a nature very different from other `standard' assets. Banks need cash to carry out their daily operations. However, with market polarization of roles, a handful of banks end up controlling how the liquidity provided by ECB is distributed troughout the interbank market. Hence, changes in the interbank rates alone are not a suitable indicator of the successfulness of a loose liquidity policy. 

Our results represent a starting point that can develop into fruitful avenues for future research. It would be relevant to focus on the issue of systemic risk in the interbank market, and to provide network-based measures for the risk of contagion. We could use alternative methodologies to study whether the big players can be a source of systemic risk, thus contributing to network disruptions. For this purpose, it would be relevant to study the contribution of the big players to the probability of fragmentation of a network. 

The presentation of the results in this paper has suggested a link between network formation and information. It would be important to construct formal measures of asymmetric information, such as the probability of informed trading of \citet{ohara}. We could then study the relation between asymmetric information and indicators of network structure. Since information is often argued as a determinant of asset prices \citep[e.g., see][]{ohara:1}, we can consider the joint contribution of private information and network centrality in the determination of lending rates.

\newpage

\singlespacing

\bibliographystyle{econometrica}%
\bibliography{examplebib}%

\clearpage

\clearpage

\clearpage

\clearpage

\clearpage

\clearpage

\clearpage

\clearpage

\begin{table}
\begin{center}
\caption{Descriptive statistics on the number of banks and trades}
\begin{tabular}{rrrrr}
\hline
{Year} & {N. banks} & {N. trades} & {N. ON trades} & {\% ON trades} \\ \hline
2006 & 172 & 115,886  & 90,368 & 77.79\% \\
2007 & 170 &  109,178 & 86,447 & 79.18\%  \\ 
2008 & 157 &  90,534   & 75,931 & 83.87\%  \\ 
2009 & 137 &  59,969   & 52,743 & 87.95\% \\ \hline
\end{tabular}
\label{tab:tab1}
\end{center}
\end{table}

\begin{table}
\begin{center}
\caption{Descriptive statistics on traded volumes}
\begin{tabular}{lrrrr}
\hline
 & {2006} & {2007} & {2008} & {2009} \\ \hline
Mean & 24.655 & 22.079 & 19.896 & 17.391 \\
Median & 15 & 10 & 10 & 10\\
St. Dev. & 31.092 & 32.486 & 30.125 & 21.185 \\
Min & 0.050 & 0.050 & 0.050 & 0.050 \\
Max & 500 & 980 & 1050 & 1000 \\
Sum & 2,228,038.150 & 1,908,744.240 & 1,510,761.430 & 917,266.410 \\ 
\hline
\end{tabular} 
\label{Amount}
\legend{Legend: This table reports traded volumes in million Euros.}
\end{center}
\end{table}

\clearpage

\begin{table}
\begin{center}
\caption{Descriptive statistics on volumes lent and borrowed per bank}
\label{bank}
\begin{tabular}{rrrr|rrr}
\hline
\multicolumn{4}{c|}{{Amount Lent}} & \multicolumn{3}{c}{{Amount Borrowed}} \\ \hline
{Year} & {\% banks }& {Mean} & {St. Dev.} &{ \% banks} & {Mean} & {St. Dev.} \\ \hline
2006 & 92\% & 14,101.51 & 24,351.13 & 89\% & 14,562.00 & 34,341.58 \\ 
2007 & 93\% & 12,079.71 & 18,689.38 & 88\% & 12,809.26 & 25,543.71 \\ 
2008 & 93\% & 556.20 & 527.72 & 89\% & 584.21 & 792.73 \\
2009 & 92\% & 379.32 & 375.44 & 84\% & 399.56 & 536.45 \\ \hline
\end{tabular}
\legend{Legend: This table reports volumes lent and borrowed in million Euros.}
\end{center}
\end{table}

\clearpage

\begin{sidewaystable}
\begin{center}
\caption{Pearson's correlation}
\label{tab:cor}
\begin{tabular}{lccc|lccc}\hline
 &  & {Incoming ties} & & & & {Outgoing ties} \\ \hline
 & {Indegree} & {Generalized Indegree} & {Instrength} && {Outdegree} & {Generalized Outdegree} & {Outstrength} \\ \hline
 &  & {2006} & & & & {2006} \\ \hline 
Indegree & 1.00 &  & &Outdegree & 1.00 &  &   \\ 
Generalized Indegree & 0.860 & 1.000 & & Generalized outdegree & 0.810 & 1.000 &  \\ 
Instrength & 0.680 & 0.950 & 1.000 & Outstrength & 0.560 & 0.920 & 1.000\\ \hline
 &  & {2007} & & & & {2007} \\ \hline
Indegree & 1.000 &  & &Outdegree & 1.000 &  &  \\ 
Generalized Indegree & 0.900 & 1.000 &  &Generalized outdegree & 0.810 & 1.000 &\\ 
Instrength & 0.750 & 0.960 & 1.000 & Outstrength & 0.610 & 0.950 & 1.000\\ \hline
 &  & {2008} & & & & {2008} \\ \hline
Indegree & 1.000 &  &  &Outdegree & 1.000 &  &  \\ 
Generalized Indegree & 0.880 & 1.000 &  &Generalized outdegree & 0.810 & 1.000 &\\ 
Instrength & 0.740 & 0.960 & 1.000 &Outstrength & 0.530 & 0.910 & 1.000 \\ \hline
 &  & {2009} & & & & {2009} \\ \hline
Indegree & 1.000 &  & &Outdegree & 1.000 &  &  \\ 
Generalized Indegree & 0.840 & 1.000 & &Generalized outdegree & 0.850 & 1.000 & \\ 
Instrength & 0.620 & 0.930 & 1.000& Outstrength & 0.650 & 0.940 & 1.000 \\ \hline
\end{tabular}

\changecaptionwidth{}\legend{Note: The correlation coefficients are all significant at 0.05 level.}
\end{center}
\end{sidewaystable}

\clearpage

																		\begin{sidewaystable}
\begin{center}
\caption{Descriptive statistics on lending activities by the big providers to the whole market}
\begin{tabular}{lrrrrrrrrrr}
\hline
	&	Market 2006	&	IT0187	&	IT0193	&	IT0258	&	IT0259	&	IT0261	&	IT0265	& IT0269 &	IT0279	&	GR0006	\\ \hline
Rate	&	2.82 (0.384)	&	2.789	&	2.864	&	2.706	&	2.768	&	2.777	&	2.870	& 2.787 &	2.915	&	2.804	\\
Std.	&		&	0.401	&	0.394	&	0.328	&	0.342	&	0.350	&	0.396	& 0.360 & 	0.358	&	0.416	\\
Tot. amount lent	&	2228038	&	52539	&	35522	&	81526	&	67074	&	67559	&	74687	&198224 &	130185	&	46649	\\
Perc.	&	100.000	&	2.358	&	1.594	&	3.659	&	3.010	&	3.032	&	3.352	& 8.897& 	5.843	&	2.094	\\
N. trades	&	90368	&	3394	&	2852	&	2975	&	2741	&	1497	&	954	& 5397&	4491	&	934	\\ \hline
	&		&		&		&		&		&		&		&		&		\\
	&	Market 2007	&	IT0175	&	IT0187	&	IT0193	&	IT0224	&	IT0257	&	IT0259 & IT0261	&	IT0269	&	IT0279	\\ \hline
Rate	&	3.874 (0.252)	&	3.857	&	3.889	&	3.896	&	3.913	&	3.775	&	3.803	&3.950&	3.870	&	3.811	\\
Std.	&		&	0.256	&	0.238	&	0.258	&	0.271	&	0.208	&	0.246	&0.210&	0.254	&	0.243	\\
Tot. amount lent	&	1908744	&	32997	&	61172	&	39924	&	39444	&	99478	&	62624	&114079 &	106468	&	77576	\\
Perc.	&	100.000	&	1.729	&	3.205	&	2.092	&	2.066	&	5.212	&	3.281	&5.977&	5.578	&	4.064	\\
N. trades	&	86447	&	2352	&	4320	&	3633	&	1723	&	2961	&	936	& 2936&	4005	&	3538	\\ \hline
	&		&		&		&		&		&		&		&		&		\\
	&	Market 2008	&	IT0173	&	IT0186	&	IT0211	&	IT0224	&	IT0247	&	IT0255 & IT0264	&	IT0269	&		\\ \hline
Rate	&	3.874 (0.252)	&	3.799	&	3.930	&	3.823	&	3.938	&	3.695	&	3.919	&3.948&	3.852	&		\\
Std.	&		&	0.579	&	0.462	&	0.499	&	0.440	&	0.549	&	0.475	&0.386&	0.438	&		\\
Tot. amount lent	&	1510761	&	25633	&	45006	&	17466	&	66864	&	32318	&	88875	&56106&	48137	&		\\
Perc.	&	100.000	&	1.697	&	2.979	&	1.156	&	4.426	&	2.139	&	5.883	&3.714&	3.186	&		\\
N. trades	&	75931	&	1710	&	1633	&	2272	&	2479	&	854	&	1674	& 995 &	1852	&		\\ \hline
	&		&		&		&		&		&		&		&		&		\\
	&	Market 2009	&	IT0173	&	IT0175	&	IT0186	&	IT0197	&	IT0209	&	IT0224	& IT0278		&		\\ \hline
Rate	&	0.66(0.45)	&	0.729	&	0.616	&	0.672	&	0.468	&	0.617	&	0.576	&	0.761	&		\\
Std.	&		&	0.520	&	0.356	&	0.365	&	0.327	&	0.476	&	0.367	&	0,510	&		\\
Tot. amount lent	&	917266	&	23492	&	20170	&	22248	&	39087	&	8502	&	48730	&25744		&		\\
Perc.	&	100.000	&	2.561	&	2.199	&	2.425	&	4.261	&	0.927	&	5.312	&	2.807	&		\\
N. trades	&	52743	&	1121	&	1396	&	668	&	1571	&	1445	&	2018	&	1153	&		\\ \hline
\end{tabular}\normalsize
\label{lending_big_providers_to_market}

\changecaptionwidth{}\legend{Legend: Rate: mean rate; Std.: standard deviation of mean rate; Tot. amount lent: total amount lent; Perc.: market share of total amount traded; N. trades: number of trades.}
\end{center}
\end{sidewaystable}

\clearpage

																		\begin{sidewaystable}
\begin{center}
\caption{Descriptive statistics on lending activities by the big losers to the whole market}
\begin{tabular}{lrrrrrrrrrr}
\hline
	&	Market 2006	&	IT0162	&	IT0180	&	IT0203	&	IT0210	&	IT0267	&	IT0268	&	IT0270	&	IT0271	&	IT0272	\\ \hline
Rate	&	2.82 (0.384)	&	3.013	&	2.575	&	2.888	&	2.883	&	3.022	&	2.938	&	2.971	&	2.854	&	2.757	\\
Std.	&		&	0.356	&	0.175	&	0.348	&	0.416	&	0.449	&	0.460	&	0.559	&	0.378	&	0.334	\\
Tot. amount lent	&	2228038	&	1908	&	100	&	7130	&	113	&	2386	&	1600	&	8353	&	84537	&	4463	\\
Perc.	&	100.000	&	0.086	&	0.004	&	0.320	&	0.051	&	0.107	&	0.072	&	0.375	&	3.794	&	0.200	\\
N. trades	&	90368	&	137	&	5	&	216	&	68	&	177	&	146	&	214	&	962	&	165	\\ \hline
	&		&		&		&		&		&		&		&		&		&		\\
	&	Market 2007	&	IT0162	&	IT0165	&	IT0168	&	IT0203	&	IT0210	&	IT0267	&	IT0270	&	IT0272	&	IT0278	\\ \hline
Rate	&	3.874 (0.252)	&	3.830	&	3.840	&	3.965	&	3.941	&	3.523	&	3.829	&	3.830	&	3.970	&	3.876	\\
Std.	&		&	0.183	&	0.257	&	0.191	&	0.232	&	0.496	&	0.270	&	0.277	&	0.241	&	0.267	\\
Tot. amount lent	&	1908744	&	3062	&	6169	&	14918	&	20673	&	753	&	7228	&	22427	&	5142	&	9138	\\
Perc.	&	100	&	0.160	&	0.323	&	0.782	&	1.083	&	0.039	&	0.379	&	1.175	&	0.269	&	0.479	\\
N. trades	&	86447	&	253	&	166	&	87	&	666	&	62	&	475	&	602	&	213	&	656	\\ \hline
	&		&		&		&		&		&		&		&		&		&		\\
	&	Market 2008	&	IT0159	&	IT0160	&	IT0165	&	IT0168	&	IT0210	&	IT0253	&	IT0258	&	IT0278	&		\\ \hline
Rate	&	3.874 (0.252)	&	4.080	&	4.146	&	4.078	&	3.647	&	4.047	&	4.071	&	3.931	&	3.311	&		\\
Std.	&		&	0.211	&	0.203	&	0.178	&	0.459	&	0.261	&	0.188	&	0.357	&	0.689	&		\\
Tot. amount lent	&	1510761	&	18752	&	11634	&	202	&	2780	&	3862	&	1199	&	6438	&	10126	&		\\
Perc.	&	100	&	1.241	&	0.770	&	0.013	&	0.184	&	0.256	&	0.079	&	0.426	&	0.670	&		\\
N. trades	&	75931	&	336	&	327	&	19	&	88	&	216	&	10	&	255	&	594	&		\\ \hline
	&		&		&		&		&		&		&		&		&		&		\\
	&	Market 2009	&	IT0159	&	IT0165	&	IT0168	&	IT0253	&	IT0265	&	IT0284	&		&		&		\\ \hline
Rate	&	0.66(0.45)	&	0.389	&	0.282	&	0.651	&	0.370	&	0.838	&	0.632	&		&		&		\\
Std.	&		&	0.184	&	0.057	&	0.557	&	0.226	&	0.485	&	0.434	&		&		&		\\
Tot. amount lent	&	917266	&	3510	&	735	&	124	&	5304	&	20444	&	13525	&		&		&		\\
Perc.	&	100	&	0.383	&	0.080	&	0.014	&	0.578	&	2.229	&	1.475	&		&		&		\\
N. trades	&	52743	&	29	&	40	&	8	&	27	&	537	&	352	&		&		&		\\ \hline
\end{tabular}\normalsize
\label{lending_big_losers_to_market}

\changecaptionwidth{}\legend{Legend: Rate: mean rate; Std.: standard deviation of mean rate; Tot. amount lent: total amount lent; Perc.: market share of total amount traded; N. trades: number of trades.}
\end{center}
\end{sidewaystable}

\clearpage

																		\begin{sidewaystable}
\begin{center}
\caption{Descriptive statistics on borrowing activities by the big providers from the whole market}
\begin{tabular}{lrrrrrrrrrr}\hline			
	&	Market 2006	&	IT0187	&	IT0193	&	IT0258	&	IT0259	&	IT0261	&	IT0265	&	IT0269	&	IT0279	&	GR0006	\\ \hline
Rate	&	2.82 	&	0	&	0	&	3.135	&	3.171	&	2.795	&	2.535	&	2.968	&	2.759	&	2.849	\\
Std.	&		&	0	&	0	&	0.314	&	0.391	&	0.422	&	0.404	&	0.451	&	0.178	&	0.095	\\
Tot. amount borrowed &	2228038	&	0	&	0	&	5361	&	2075	&	13960	&	5301	&	2046	&	39	&	563	\\
Perc.	&	100	&	0	&	0	&	0.241	&	0.093	&	0.627	&	0.238	&	0.092	&	0.002	&	0.025	\\
N. trades	&	90368	&	0	&	0	&	293	&	105	&	544	&	260	&	110	&	6	&	11	\\ \hline
	&		&		&		&		&		&		&		&		&		&		\\

	&	Market 2007	&	IT0175	&	IT0187	&	IT0193	&	IT0224	&	IT0257	&	IT0259	&	IT0261	&	IT0269	&	IT0279	\\ \hline
Rate	&	3.874 	&	0	&	0	&	3.807	&	3.811&	3.952	&	3.941	&	3.873	&	4.029	&	3.827	\\
Std.	&		&	0	&	0	&	0.434	&	0.113	&	0.312	&	0.225	&	0.224	&	0.252	&	0.255	\\
Tot. amount borrowed &	1908744	&	0	&	0	&	5	&	277	&	9157	&	7052	&	11178	&	964	&	276	\\
Perc.	&	100	&	0	&	0	&	0.0002	&	0.015	&	0.480	&	0.369	&	0.586	&	0.051	&	0.014	\\
N. trades	&	86447	&	0	&	0	&	2	&	30	&	186	&	234	&	409	&	40	&	24	\\ \hline
	&		&		&		&		&		&		&		&		&		&		\\

	&	Market 2008	&	IT0173	&	IT0186	&	IT0211	&	IT0224	&	IT0247	&	IT0255	&	IT0264	&	IT0269	&		\\ \hline
Rate	&	3.874 	&	3.4	&	4.075	&	3.907	&	4.110	&	3.165	&	3.962	&	4.07	&	3.958	&		\\
Std.	&		&	0.141	&	0.084	&	0.821	&	0.218	&	1.364	&	0.241	&		&	0.236	&		\\
Tot. amount borrowed &	1510761	&	20	&	267	&	35	&	191	&	19	&	7956	&	11	&	226	&		\\
Perc.	&	100	&	0.001	&	0.018	&	0.002	&	0.013	&	0.001	&	0.527	&	0.001	&	0.015	&		\\
N. trades	&	75931	&	2	&	26	&	7	&	13	&	2	&	127	&	1	&	21	&		\\ \hline
	&		&		&		&		&		&		&		&		&		&		\\

	&	Market 2009	&	IT0173	&	IT0175	&	IT0186	&	IT0197	&	IT0209	&	IT0224	&	IT0278	&		&		\\ \hline
Rate	&	0.66	&	0	&	0	&	1.102	&	1.316	&	0.703	&	1.3	&	0.536	&		&		\\
Std.	&		&	0	&	0	&	0.046	&	0.788	&	0.328	&		&	0.421	&		&		\\
Tot. amount borrowed &	917266	&	0	&	0	&	106	&	1645	&	61	&	30	&	5503	&		&		\\
Perc.	&	100	&	0	&	0	&	0.012	&	0.179	&	0.007	&	0.003	&	0.600	&		&		\\
N. trades	&	52743	&	0	&	0	&	14	&	117	&	12	&	1	&	377	&		&		\\ \hline

\end{tabular}\normalsize
\label{borrowing_big_providers}

\changecaptionwidth{}\legend{Legend: Rate: mean rate; Std.: standard deviation of mean rate; Tot. amount lent: total amount borrowed; Perc.: market share of total amount traded; N. trades: number of trades.}
\end{center}
\end{sidewaystable}

\clearpage

																		\begin{sidewaystable}
\begin{center}
\caption{Descriptive statistics on borrowing activities by the big losers from the whole market}
\begin{tabular}{lrrrrrrrrrr}
\hline	
	&	Market 2006	&	IT0162	&	IT0180	&	IT0203	&	IT0210	&	IT0267	&	IT0268	&	IT0270	&	IT0271	&	IT0272	\\ \hline
Rate	&	2.82 	&	2.753	&	2.455	&	2.858	&	2.874	&	2.752	&	2.830	&	2.804	&	2.801	&	2.850	\\
Std.	&		&	0.382	&	0.137	&	0.414	&	0.380	&	0.354	&	0.366	&	0.323	&	0.372	&	0.393	\\
Tot. amount borrowed	&	2228038	&	78235	&	71569	&	81035	&	119242	&	109766	&	143001	&	146675	&	211838	&	224535	\\
Perc.	&	100	&	3.511	&	3.212	&	3.637	&	5.352	&	4.927	&	6.418	&	6.583	&	9.508	&	10.078	\\
N. trades	&	90368	&	4448	&	1263	&	2497	&	6410	&	4480	&	6209	&	6631	&	4515	&	6931	\\ \hline
	&		&		&		&		&		&		&		&		&		&		\\
	&	Market 2007	&	IT0162	&	IT0165	&	IT0168	&	IT0203	&	IT0210	&	IT0267	&	IT0270	&	IT0272	&	IT0278	\\ \hline
Rate	&	3.874 	&	3.912	&	3.906	&	3.769	&	3.814	&	3.905	&	3.822	&	3.911	&	3.813	&	3.873	\\
Std.	&		&	0.245	&	0.225	&	0.247	&	0.264	&	0.239	&	0.314	&	0.219	&	0.222	&	0.231	\\
Tot. amount borrowed	&	1908744	&	82588	&	75635	&	40668	&	60414	&	123144	&	66725	&	95608	&	183844	&	82050	\\
Perc.	&	100	&	3.707	&	3.395	&	2.131	&	3.165	&	6.452	&	3.496	&	5.009	&	9.632	&	4.299	\\
N. trades	&	86447	&	4129	&	4340	&	1397	&	2063	&	5643	&	3199	&	4246	&	5252	&	4017	\\ \hline
	&		&		&		&		&		&		&		&		&		&		\\
	&	Market 2008	&	IT0159	&	IT0160	&	IT0165	&	IT0168	&	IT0210	&	IT0253	&	IT0258	&	IT0278	&		\\ \hline
Rate	&	3.874 	&	3.787	&	3.669 &	3.855	&	3.931	&	3.982	&	3.810	&	3.919	&	4.036	&		\\
Std.	&		&	0.490	&	0.609	&	0.515	&	0.510	&	0.394	&	0.424	&	0.4547	&	0.371	&		\\
Tot. amount borrowed	&	1510761	&	101169	&	63802	&	84500	&	51164	&	37799	&	85240	&	75973	&	115952	&		\\
Perc.	&	100	&	6.697	&	4.223	&	5.593	&	3.387	&	2.502	&	5.642	&	5.029	&	7.675	&		\\
N. trades	&	75931	&	2661	&	2568	&	5419	&	1835	&	1906	&	1883	&	3549	&	5346	&		\\ \hline
	&		&		&		&		&		&		&		&		&		&		\\
	&	Market 2009	&	IT0159	&	IT0165	&	IT0168	&	IT0253	&	IT0265	&	IT0284	&		&		&		\\ \hline
Rate	&	0.66	&	0.503	&	0.746	&	0.5357	&	0.492	&	0.632	&	0.610	&		&		&		\\
Std.	&		&	0.430&	0.489&	0.329	&	0.383&	0.426&	0.360
	&		&		&		\\
Tot. amount borrowed	&	917266	&	49793	&	47940	&	130187	&	27500	&	74518	&	70486	&		&		&		\\
Perc.	&	100	&	2.000	&	5.226	&	14.193	&	2.998
	&	8.124	&	7.684	&		&		&		\\
N. trades	&	52743	&	1755	&	3291	&	5552	&	734	&	3288	&	2977	&		&		&		\\ \hline

\end{tabular}\normalsize
\label{borrowing_big_losers}

\changecaptionwidth{}\legend{Legend: Rate: mean rate; Std.: standard deviation of mean rate; Tot. amount: total amount borrowed; Perc.: market share of total amount traded; N. trades: number of trades.}
\end{center}
\end{sidewaystable}

\end{document}